\documentclass[12pt]{iopart}
\pdfoutput=1
\usepackage{iopams}  
\usepackage{cite}
\usepackage{graphicx}
\usepackage[colorlinks]{hyperref}

\def\mathhyphen{{\hbox{-}}}

\begin{document}

\title[Statistical mechanics of complex economies]{Statistical mechanics of complex economies}

\author{Marco Bardoscia\footnote{Any views expressed are solely those of the author(s) and so cannot be taken to represent those of the Bank of England or to state Bank of England policy.}}
\address{Bank of England, 20 Moorgate, London EC2R 6DA, UK}
\ead{marco.bardoscia@bankofengland.co.uk}

\author{Giacomo Livan}
\address{Department of Computer Science, University College London, Gower St., London WC1E 6BT, UK}
\address{Systemic Risk Centre, London School of Economics and Political Sciences, Houghton St., London WC2A 2AE, UK}
\ead{g.livan@ucl.ac.uk}

\author{Matteo Marsili}
\address{Abdus Salam International Centre for Theoretical Physics, Strada Costiera 11, 34151 Trieste, Italy}
\ead{marsili@ictp.it}

\begin{abstract}
In the pursuit of ever increasing efficiency and growth, our economies have evolved to remarkable degrees of complexity, with nested production processes feeding each other in order to create products of greater sophistication from less sophisticated ones, down to raw materials. The engine of such an expansion have been competitive markets that, according to General Equilibrium Theory (GET), achieve efficient allocations under specific conditions. We study large random economies within the GET framework, as templates of complex economies, and we find that a non-trivial phase transition occurs: the economy freezes in a state where all production processes collapse when either the number of primary goods or the number of available technologies fall below a critical threshold. As in other examples of phase transitions in large random systems, this is an {\em unintended consequence} of the growth in complexity. Our findings suggest that the Industrial Revolution can be regarded as a sharp transition between different phases, but also imply that well developed economies can collapse if too many intermediate goods are introduced.
\end{abstract}

%
%
%
\maketitle
%
%

\section{Introduction}
Complex artefacts, such as a computer or a car, involve a large number of components, each of which is the result of a different production process, requiring inputs from yet other production processes. Many of these processes, once internal to the same firm, are typically delocalised in a myriad of firms whose interaction is mediated through market prices \cite{Langlois}. The traditional method to parse the production activity of an economy is Input-Output (IO) economic analysis, which decomposes an economy's productive sector into the elementary flows between its components.
In its original inception \cite{Leontief} this kind of analysis simply amounted to compiling and inverting IO matrices to determine the amount of goods to be produced by each sector in order for the economy to match consumer demand. Modern IO analysis is instead carried out in the framework of economics' General Equilibrium Theory (GET), which seeks to derive macro-economic behaviour from the interaction between profit-maximizing firms and utility-maximizing consumers through market prices.

Attempts to derive laws for the collective behaviour of large economies are hampered by the dazzling complexity of the diverse agents involved and of their network of interactions. These difficulties are only partly circumvented by the so-called {\em representative agent} approach, that effectively derives macro-economic dynamics by scaling up insights on micro-economics. This approach provides intuition on a plethora of phenomena, and is the basis of the most elaborate computational Dynamic Stochastic General Equilibrium (DSGE) models that are used in practice for policy analysis \cite{DSGE}. Yet, its conceptual \cite{Kirman} and practical \cite{ChangKimSchorfheide} shortcomings have been repeatedly pointed out.

On the one hand, efforts in GET have focused in trying to prove results for economies under the broadest possible assumptions. This program is similar to worst-case analysis in computer science, where a problem complexity is determined by estimating how hard it is to solve it in the hardest possible instance. Yet, worst-case instances are often very different from typical ones, which is why this approach has lately been contrasted with typical-case complexity. Likewise, even if no result can be proved for all economies, the typical behavior of large economies may be well defined and it can be characterized with the techniques described in this paper. We believe that the GET program, which was abandoned in its ``worst-case'' version, can be revamped in its ``typical-case'' version.

On the other hand, GET aimed at a description of an economy which is similar to the one that Newton's classical mechanics provides of a gas of particles. While being extremely detailed, it is hardly of any use for describing the behavior of a gas. Yet, when classical mechanics is combined with a statistical ansatz on the distribution of micro-states, a full description of the typical properties of gases that quantitatively reproduces the laws of thermodynamics emerges. Here we carry out a similar program, by deriving a description of the typical properties of an economy that enjoys many of the properties of such an approach in physics, such as the irrelevance of the micro-economic details on the collective behavior.

Statistical mechanics \cite{Chandler} offers a general perspective \cite{Jaynes} on collective phenomena, which also takes the heterogeneity at the micro scale fully into account. This has shed considerable light on the physics of disordered systems, such as glasses and random alloys \cite{MezardParisiVirasoro}, but also on a wide variety of subjects, including atomic spectra of heavy ions \cite{Wigner}, the stability of ecosystems \cite{May}, algorithmic transitions in computer science \cite{MZKST,MPZ}, statistics \cite{MarchenkoPastur}, random geometries \cite{donoho-tanner}, and portfolio instability in finance \cite{Kondor}. One of the main insights provided by this approach (and common to the aforementioned examples) is the presence of potential sharp changes in the collective behaviour -- called phase transitions -- that can hardly be explained from the behaviour of individual components.

The purpose of the present paper is to show that an unexpected sharp phase transition can also plague large complex economies as described by GET. Specifically, we show that when the fraction of non-primary goods, i.e.\ goods that result as an output of a production process, exceeds a critical threshold, the economy freezes in a state where all production processes collapse. This transition is reminiscent of the one discussed by Donoho and Tanner \cite{donoho-tanner} for random geometries in high dimensions, although it is of a slightly different nature. The occurrence of the transition only depends on the properties of the production set, and is independent of the properties of consumers. Within the simplified description of an economy that GET provides, this result not only suggests a sharp separation between industrialised and underdeveloped economies, but it also implies a collapse of the economy when the number of intermediate goods grows too large.


\section{The General Equilibrium framework}
We consider a classical model \cite{Lancaster} of an economy composed of $C$ goods, where producers aim for the maximum profit and consumers adjust their demand in order to maximize their own utility. Both producers and consumer maximisation problems are solved at fixed prices, that are tuned to match demand and supply for each good in competitive markets. This is a single period model, where the consumer is provided with a basket of primary goods that are then exchanged in markets and transformed into final consumption goods by competitive firms in the production sector. It can be helpful to think of the primary goods as the goods that are readily available in Nature and that do not need to be produced. Therefore, such goods constitute the initial endowment with which the consumer is equipped. The emphasis is on the ability of the production sector to transform the abundant primary goods into the desired (scarce) final goods. A key aspect is that {\em primary goods} may be different from {\em final goods}, and some of the goods may be neither primary nor final. These are {\em intermediate goods}, that enter in the transformation process of primary goods into final ones.

Formally, let {$x_0^c$ be the aggregate initial endowment of the consumer for good $c$, with $c=1,\ldots,C$, which is strictly positive for primary goods and equal to zero for non-primary goods. Hence, $\mathcal{P} = \{c : x_0^c > 0\}$ is the set of \emph{primary} goods, and $x_0^c=0$ for all $c\not\in\mathcal{P}$. Likewise, final goods $c\in \mathcal{F}$ are those the utility of consumers depends on.
At fixed prices $\boldsymbol{p}$, the consumer's problem is that of exchanging the initial endowments
for the final consumption goods, in order to achieve utility maximization compatibly with a budget constraint.

The role of the production sector is to transform initial endowments 
into final goods.
We assume a linear activity model for the production sector, with $N$ transformation processes, each characterised by a vector $\boldsymbol{q}$. Good $c$ is an input (output) if $q^c$, the $c^{\rm th}$ component of $\boldsymbol{q}$, is negative (positive).
When such an activity operates at scale $s_i \ge 0$, the amount of good $c$ consumed or produced by the $i^{\rm th}$ activity is simply $s_i q_i^c$. So the whole production sector is defined by $N$ vectors $\boldsymbol{q}_i$, $i=1,\ldots,N$, each run at a scale $s_i\ge 0$. As customary \cite{Lancaster}, we also assume disposal technologies for each good\footnote{These corresponds to vectors $d_c$ with all components equal to zero, apart from component $c$, $d_c^c=-1$. Disposal technologies are not included in equation (\ref{eq:market_clearing}) with the understanding that those goods with $x^c>0$ that are not final are disposed of, i.e.\ they are wasted.}. The {\em feasible set} of production scales $\{s_i\}$ is given by all vectors $\boldsymbol{s}=(s_1,\ldots,s_N)$ such that
\numparts
\begin{eqnarray}
&x^c=x^c_0+\sum_{i=1}^Ns_iq_i^c \label{eq:market_clearing} \\
&x^c \ge 0 \, ,\qquad \forall c=1,\ldots,C \, . \label{eq:feasible}
\end{eqnarray}
\endnumparts
As we can see from equation (\ref{eq:market_clearing}), $x^c$ is the sum of the initial endowments $x_0^c$ and of the aggregate net production of good $c$, and it can be therefore interpreted as the available volume of good $c$.
We now have two possibilities: either good $c$ is final, or it is not. 
In the former case, by assuming strongly monotonic preferences, consumers will always consume all the available volume of good $c$. 
Hence, for final goods, $x^c$ is equal to the level of consumption of good $c$.
In the latter case, consumers will not consume any volume of good $c$, and therefore $x^c$ is equal to the excess supply of good $c$.
Equation (\ref{eq:market_clearing}) implies that excess supply cannot be negative for any good.

At market prices $\boldsymbol{p}$, the profit of each activity $i$ when run at scale $s_i$ is given by $s_i\boldsymbol{p}\cdot\boldsymbol{q}_i$. At equilibrium, each $s_i$ is fixed within the feasible set in order to maximise profit. If a technology is unprofitable with the equilibrium prices, it is optimal to stop operating it by setting its scale to zero. Hence, the number $N_>=|\{i:~s_i>0\}|$ of activities that actually operate will be smaller than $N$ in general.
Finally, prices are set so as to match supply with consumer demand.
Two important generic properties of the equilibrium can be easily derived by multiplying both sides of equation (\ref{eq:market_clearing}) by $p^c$ (the price of commodity $c$) and by summing over all commodities. One immediately finds that consumers saturate their budget constraint, i.e.\ $\boldsymbol{p}\cdot\boldsymbol{x}=\boldsymbol{p}\cdot\boldsymbol{x}_0$ (Walras' law), and that the profit for each activity is zero.
The number of active production processes is at most $N_>\le C$.
Final goods turn out to be associated with positive prices, whereas for the remaining goods, prices are set by the marginal profits \cite{Lancaster}.
Non-final goods that are in positive excess supply $x^c>0$ have $p^c=0$ and we will interpret them as \emph{waste}.
Among these, goods that are also not primary (i.e.\ $c\in\overline{\mathcal{F}}\bigcap\overline{\mathcal{P}}$) are intermediate goods; those among them that have $x^c=0$ are fully exploited by activities, and therefore have $p^c>0$.
Besides these generic results, little can be said about how the properties of an economy (such as levels of consumption, scales of production, or fraction of operating activities) depend on its structure, i.e.\ on the number of goods of different types, on the number of technologies, etc. Indeed the project of general equilibrium was largely abandoned for its lack of specific predictions \cite{SMD30}.


\section{Large random economies} \label{sec:random}
Rather than considering a specific realisation of the framework discussed above, we discuss an ensemble of economies drawn from a given distribution. The key observation is that, when the economy becomes large enough, certain properties -- called {\em self-averaging} -- exhibit the same collective behaviour for almost all realisations. In view of their statistical robustness, these properties are the natural candidates to be compared to the observed aggregate behaviour of complex systems, an approach that has been remarkably successful in a variety of contexts \cite{Wigner,MZKST,MezardParisiVirasoro}, including systems of heterogeneous interacting agents \cite{DeMartinoMarsili}.

Our main results only entail properties of the production sector, so we shall avoid the intricacies of the aggregation problem on the demand side, and specialise to the simpler case of one (representative) consumer with a separable utility function
\begin{equation} \label{eq:utility}
U(\boldsymbol{x})=\sum_{c\in\mathcal{F}} u(x^c) = \sum_{c=1}^C k^c u(x^c) \ ,
\end{equation}
where $u(\cdot)$ is a concave increasing function (i.e.\ $u^\prime>0$ and $u^{\prime \prime}<0$) 
and $\boldsymbol{k}$ encodes consumer preferences, as $k^c = 1$ ($k^c = 0$) if the good $c$ is final (non-final).
As discussed above, the utility function only depends on the {\em final} goods $c\in\mathcal{F}$. Each good is assigned  to the class $\mathcal{F}$ with probability $f$ and, independently, to class $\mathcal{P}$ with probability $\pi$. So the number of final (primary) goods is $|\mathcal{F}|=fC$ ($|\mathcal{P}|=\pi C$). A primary good $c$ is part of the initial endowments, and therefore $x_0^c=1$, whereas $x_0^c=0$ for all non-primary goods.
This fully specifies the demand side of the economy.

As for the production sector, we take a maximum entropy approach in the spirit of Ref.\ \cite{Jaynes}, where the only assumption we make is that the first two moments $\boldsymbol{q}_i\cdot\boldsymbol{1}=\sum_{c=1}^C q_i^c=-\epsilon$ and $\boldsymbol{q}_i\cdot\boldsymbol{q}_i=1$ are fixed. This implies that each activity $\boldsymbol{q}_i$ is an independently drawn random vector satisfying these constraints. Here $\epsilon>0$ means that, for each technology, the quantity of inputs is larger than the quantity of outputs. This ensures that no linear combination of the activities with non-negative coefficients $s_i$ can produce some output without any input. Therefore, $\epsilon>0$ encodes irreversibility and its value is a measure of the inefficiency of production processes.

%
%

The convexity of $U$ ensures that the equilibrium is unique \cite{Lancaster} and it satisfies the First Welfare theorem. This can be rephrased by saying that, when the market clears, the optimal production scales $\boldsymbol{s}^*=(s_1^*,\ldots,s_N^*)$ deliver an optimal consumption bundle $\boldsymbol{x}^*$ to consumers, given by the market clearing condition equation (\ref{eq:market_clearing}). 
From the perspective of the optimization problem, the aforementioned constraint $\boldsymbol{p}\cdot\boldsymbol{x}=\boldsymbol{p}\cdot\boldsymbol{x}_0$ is accounted for simply by substituting equation (\ref{eq:market_clearing}) in equation (\ref{eq:utility}).
This implies that (see \cite{DeMartinoMarsiliPeresCastillo}) the equilibrium is given by the solution of
\begin{equation} \label{eq:max}
\max_{\boldsymbol{s}\geq 0} U\left(\boldsymbol{x}_0+\sum_{i=1}^Ns_i\boldsymbol{q}_i\right).
\end{equation}
All other properties of the equilibrium can be computed from the solution $\boldsymbol{s}^*$ and the market clearing condition.
The solution $\boldsymbol{s}^*$  depends on the specific (random) realisation of the economy, i.e.\ precisely on which goods are final and/or primary and on the specific realisation of the activities. Yet, if the economy is large enough, i.e.\ for large values of $N$ and $C$, the aforementioned self-averaging quantities attain {\em typical} values with very high probability, and independently of the specific realisation.
For example, the average scale of production $\langle s^*\rangle = \frac{1}{N}\sum_i s_i^*$ or the number of activities with $s_i^*\in[s,s+ds)$ both satisfy this property. These and other quantities can be computed analytically using techniques borrowed from statistical physics of disordered systems \cite{MezardParisiVirasoro}, which, as we shall detail in the following and in \ref{sec:sm_optm}, amounts to ``promoting'' some of the model's parameters to random variables and averaging over their probability distributions in the ``thermodynamic'' limit $N \rightarrow \infty$, $C \rightarrow \infty$, with finite ratio $n = N/C$. $n$ quantifies the number of technologies available per good, and therefore can be taken as a synthetic measure of how much an economy is developed. 
In a nutshell, by the law of large numbers, we expect that:
\begin{equation} \label{eq:utility_average}
\lim_{N \rightarrow \infty} \frac{1}{N} U(\boldsymbol{s}^* | \boldsymbol{q}, \boldsymbol{x}_0, \boldsymbol{k}) = 
\lim_{N \rightarrow \infty} \frac{1}{N} \Big\langle U(\boldsymbol{s}^* | \boldsymbol{q}, \boldsymbol{x}_0, \boldsymbol{k}) \Big\rangle_{\boldsymbol{q}, \boldsymbol{x}_0, \boldsymbol{k}} \, ,
\end{equation}
where we have made explicit the dependence of the utility function on the activities $q$, on the initial endowments $\boldsymbol{x}_0$, and on consumer preferences $\boldsymbol{k}$. 

The problem on the r.h.s.\ of equation (\ref{eq:utility_average}) entails, at least in principle, the optimization and averaging of the utility function over an infinite number of variables. Such problem can be solved analytically by resorting to the replica method \cite{MezardParisiVirasoro} borrowed from the statistical physics of disordered systems. The first step of the method consists in computing the utility function's maximum $U(\boldsymbol{s}^* | \boldsymbol{q}, \boldsymbol{x}_0, \boldsymbol{k})$ for a given realisation of the random variables $\boldsymbol{q}$, $\boldsymbol{x}_0$, and $\boldsymbol{k}$. Since the utility function is an extensive quantity (see equation (\ref{eq:utility})), for large $N$ we expect $U(\boldsymbol{s}^* | \boldsymbol{q}, \boldsymbol{x}_0, \boldsymbol{k}) / N$ to be finite. Therefore, its maximum can be computed by using the steepest descent method \cite{DeMartinoMarsiliPeresCastillo}:
\begin{equation} \label{eq:max1}
\lim_{N \rightarrow \infty} \frac{1}{N} U (\boldsymbol{s}^* | \boldsymbol{q}, \boldsymbol{x}_0, \boldsymbol{k}) = \lim_{N \rightarrow \infty} \lim_{\beta \rightarrow \infty} \frac{1}{\beta N} \log Z (\beta | \boldsymbol{q}, \boldsymbol{x}_0, \boldsymbol{k}) \ ,
\end{equation} 
where 
\begin{equation} \label{eq:partfunc}
Z (\beta | \boldsymbol{q}, \boldsymbol{x}_0, \boldsymbol{k}) = \int_0^\infty \mathrm{d} \boldsymbol{s} \ \mathrm{e}^{\beta U (\boldsymbol{s} | \boldsymbol{q}, \boldsymbol{x}_0, \boldsymbol{k})} 
\end{equation}
is called \emph{partition function} in statistical physics parlance. The rationale is that, for a large system and in the limit $\beta \to \infty$, the leading contributions to the integral in equation (\ref{eq:partfunc}) will come from the maximum of the utility function. In order to compute the right-hand side of equation (\ref{eq:utility_average}) we need to average both sides of equation (\ref{eq:max1}), and therefore we must compute the average of the logarithm of the partition function. This problem also arises in the computation of macroscopic observables of disordered systems and it can be circumvented by making use of the following identity:
\begin{equation} \label{eq:replica_trick_1}
\langle \log Z \rangle = 
\lim_{r \rightarrow 0} \frac{\langle Z^r \rangle-1}{r} \ ,
\end{equation}
so that the problem of computing $\langle \log Z \rangle$ translates into computing $\langle Z^r \rangle$, which can be done for integer values of $r$. The final step consists in performing an analytical continuation of such quantity to real values of $r$, so that the limit $r \to 0$ can be taken.

In \ref{sec:sm_optm}, we show that the approach outlined above converts the optimisation problem in equation (\ref{eq:utility_average}) into a system of six non-linear \emph{saddle-point equations}, whose unknowns are six \emph{order parameters}. Although results are derived in the limit $N\to\infty$ they provide an accurate description of the behaviour of economies for finite but large $N$ (see \ref{sec:sm_optm}). Effectively, this method reduces the optimisation problem in equation (\ref{eq:max}) to an optimisation over the parameters of a  single ``representative'' activity problem coupled to a single ``representative'' good. These features emerge from the statistical mechanical treatment, rather than being assumed at the outset as in the representative agent approach. We refer the reader to \ref{sec:sm_optm} for a detailed derivation, which follows similar lines to those in Ref.\ \cite{DeMartinoMarsiliPeresCastillo}. Let us also remark that, despite some similarities between the two frameworks, the results presented in this paper (in particular the phase transition we shall extensively discuss in the next two sections) are entirely novel.


\section{Typical properties and phase transitions} 
In summary, the parameters that define the ensemble of economies are {\em i)} the number of technologies available per good $n$, {\em ii)} the fraction $f$ of final goods, {\em iii)} the fraction $\pi$ of primary goods, {\em iv)} the inefficiency $\epsilon$ of technologies, 
and {\em v)} the utility $u(\cdot)$ of consumers. For the latter, since our qualitative results do not depend significantly on the choice of $u$ (as long as it is strictly increasing and convex), in the examples presented below we shall stick to the standard choice $u(x)=\log x$. The detailed quantitative behavior for other choices can be derived with the technique discussed in \ref{sec:sm_optm}.

In figure \ref{fig:s_phi} we plot the average optimal scale of productions $\langle s^* \rangle$ and the fraction $\phi$ of active producers as functions of $n$ and $\pi$, for a given fraction of final goods, i.e.\ for fixed preferences. We find that the parameter space is sharply divided into two regions. In the first one -- that we shall call the {\em industrial phase} -- we find a solution where a finite fraction $\phi>0$ of activities are run at positive scales $s_i^*>0$. In the second region -- the {\em pre-industrial phase} -- only solutions with $\boldsymbol{s}^*=0$ (and $\phi=0$) exist, corresponding to an economy relying exclusively on primary resources ($\boldsymbol{x}=\boldsymbol{x}_0$) with no  production activity. We shall discuss later the origin of this sharp transition. For the moment let us make two important observations. First, the value of $\pi$ at which the transition occurs is a decreasing function of $n$, meaning that 
more developed economies (with larger values of $n$) are able to sustain production with a  smaller fraction of primary goods. Second, the position of the \emph{critical line} separating the two regions does not depend on the fraction of final goods $f$, but it depends weakly on $\epsilon$. Upon decreasing $\epsilon$, i.e.\ making production processes more efficient, the critical line shifts towards lower values of $\pi$ for a fixed value of $n$, expanding the parameter space where $\langle s^* \rangle > 0$. Moreover, as $\epsilon$ decreases, $\langle s^* \rangle$ increases and gets a sharper peak around $n=2$ (see \ref{sec:sm_optm}, figure \ref{eps_peak}).

\begin{figure}
\centering
\includegraphics[width=0.49\columnwidth]{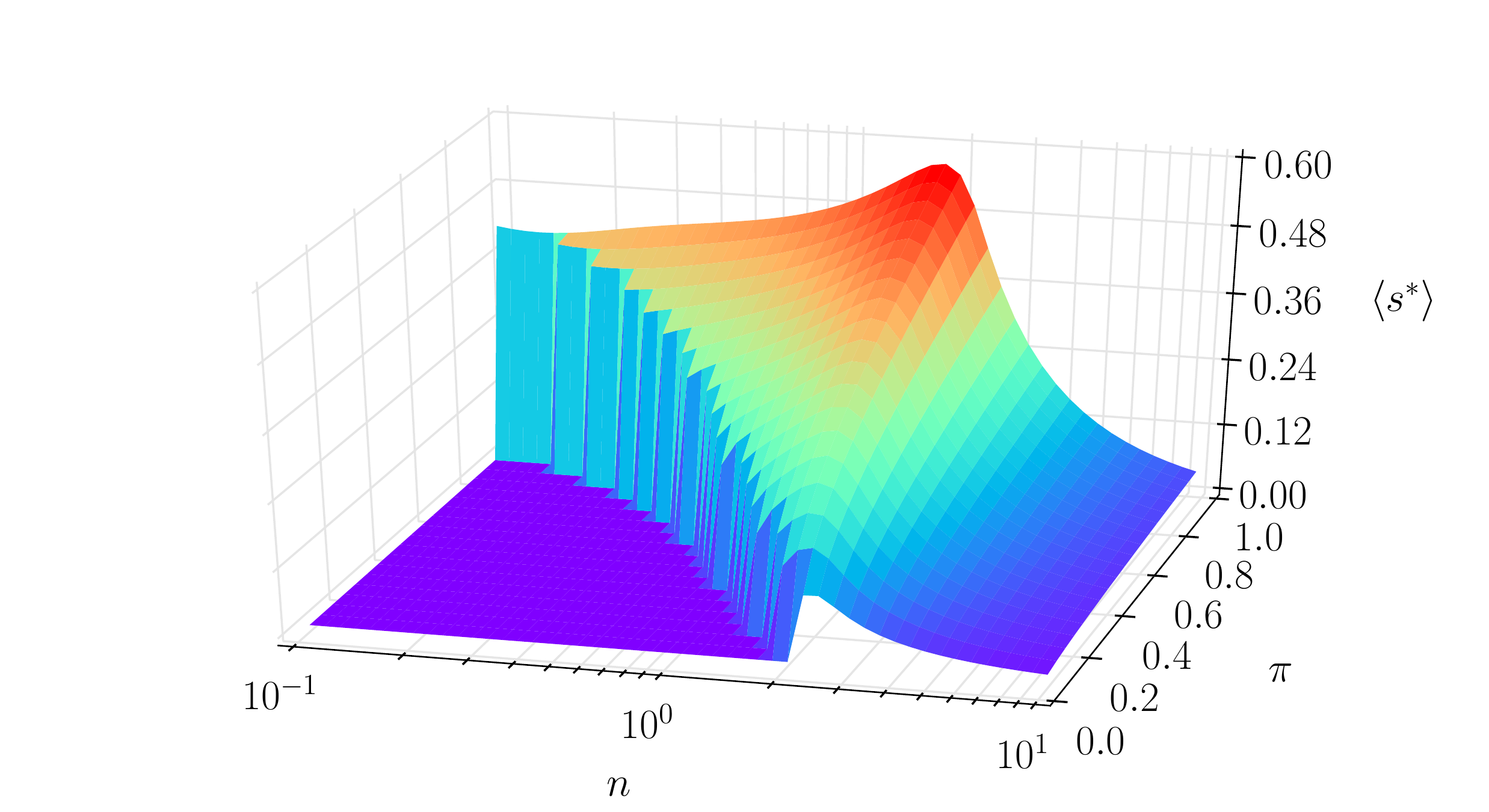}
\includegraphics[width=0.49\columnwidth]{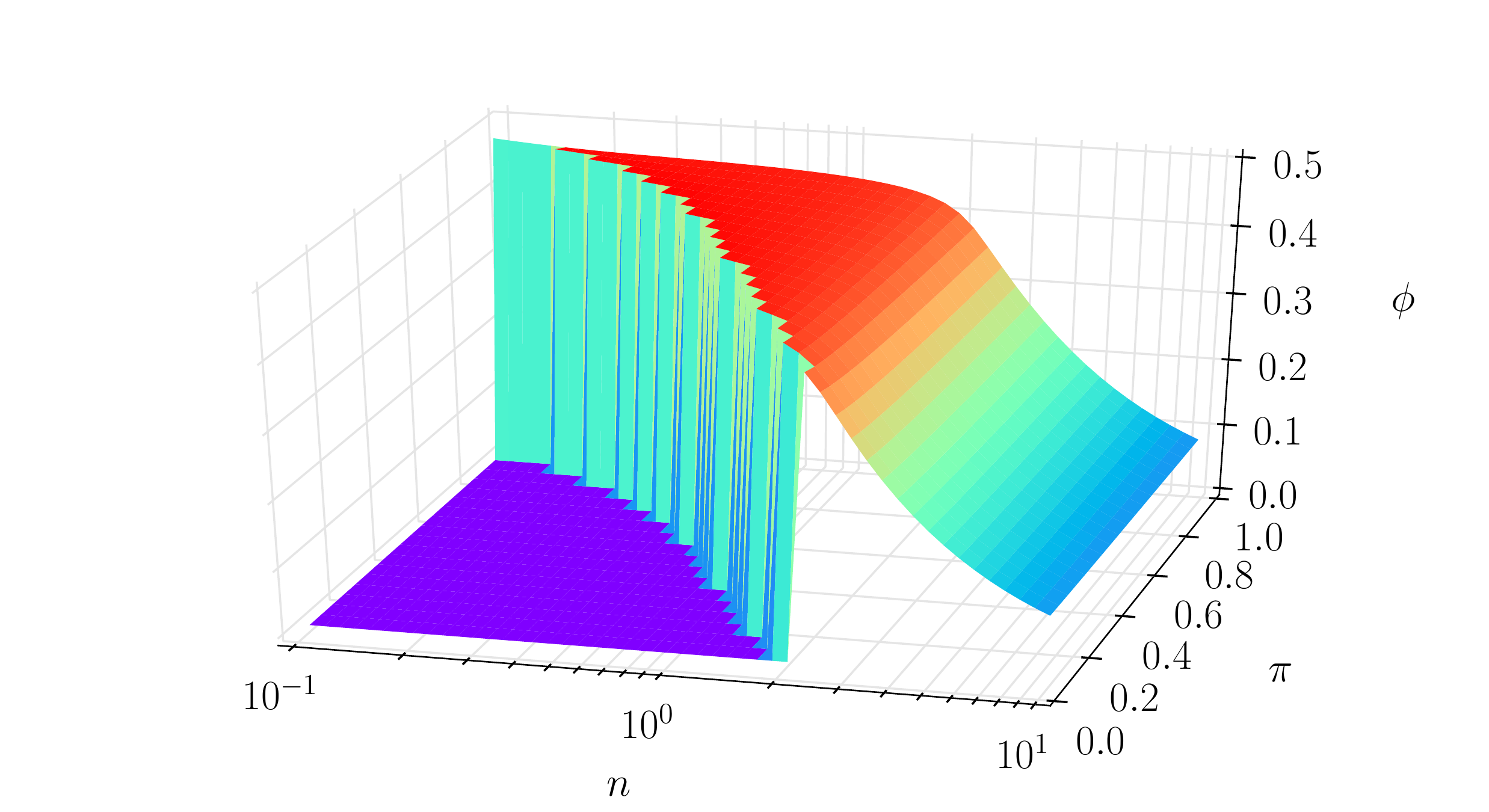}
\caption{\textbf{Regimes of the economy: production.} Optimal scales of production $\langle s^* \rangle$ (left panel) and fraction $\phi$ of active producers (right panel) for a large economy (with $f=0.5$ and $\epsilon=0.1$) as a function of the level of development of the economy $n$ and the fraction $\pi$ of primary goods. We distinguish three different regimes. The first one ($\langle s^* \rangle = \phi = 0$) describes a pre-industrial phase of the economy. The second one ($\langle s^* \rangle > 0$ and $n \lesssim 2$) described a developing stage of the economy, where the introduction of a new technology has positive spillovers on already active producers (since $\langle s^* \rangle \nearrow n$ and $\phi \nearrow n$). The third one ($\langle s^* \rangle > 0$ and $n \gtrsim 2$) describes a competitive stage, where the introduction of a new technology has disruptive effects on other ones (since $\langle s^* \rangle \searrow n$ and $\phi \searrow n$).}
\label{fig:s_phi}
\end{figure}

Let us focus on the properties of the industrial phase ($\langle s^* \rangle > 0$). As noted in \cite{DeMartinoMarsiliPeresCastillo},
we can distinguish two different regimes. For $n \lesssim 2$, the average scale of production increases with $n$, while the fraction of active producers is roughly constant. This means that introducing a new technology (by moving towards larger values of $n$) has no negative effect on the technologies already in place. In contrast, for $n \gtrsim 2$ the economy is in a highly competitive regime, in which the introduction of a new technology has a disruptive effect on the others, as both $\langle s^* \rangle$ and $\phi$ decrease with $n$.

Figure \ref{fig:wc_psi} shows the difference between these two regimes along the complementary dimension of consumption, for a given fraction of final goods, i.e.\ for fixed preferences. In order to better illustrate the economy's behavior at the transition, let us introduce the following conditional average quantities:
\begin{eqnarray} \label{x_op}
x_{11} &=& \langle x^* \rangle_{x_0=1,k=1} \qquad \mathrm{(consumed \ primary \ goods)} \\ \nonumber
x_{01} &=& \langle x^* \rangle_{x_0=0,k=1} \qquad \mathrm{(consumed \ non \mathhyphen primary \ goods)} \\ \nonumber
x_{10} &=& \langle x^* \rangle_{x_0=1,k=0} \qquad \mathrm{(wasted \ primary \ goods)} \\ \nonumber
x_{00} &=& \langle x^* \rangle_{x_0=0,k=0} \qquad \mathrm{(wasted \ non \mathhyphen primary \ goods)} \ .
\end{eqnarray}
From the above, one can introduce the average consumption $X_C$ and waste $X_W$ as
\begin{eqnarray} \label{CW}
X_C &=& f \left [\pi x_{11} + (1-\pi) x_{01} \right ] \\ \nonumber
X_W &=& (1-f) \left [\pi x_{10} + (1-\pi) x_{00} \right] \ ,
\end{eqnarray}
from which one can write $\langle x^* \rangle = X_C+X_W$ for the overall average production of goods. As can be seen in Figure \ref{fig:wc_psi}, for $n<2$ the utility and the level $X_C$ of consumption of final goods sharply increase with $n$, and, at the same time, the amount of waste $X_W$ decreases significantly. In the $n>2$ regime, instead, the utility and $X_C$ saturate to constant levels while waste $X_W$ approaches zero. Interestingly, close to $n=2$, for small $\epsilon$, a non-monotonic behaviour in $X_C$ can also occur (see figure \ref{fig:wc_psi}, bottom panel, blue solid curve and 
\ref{sec:sm_cons} for details).
As the economy exits the non-industrial phase, levels of consumption of final goods experience a jump, which can be either positive or negative depending on the inefficiency $\epsilon$. This, in turn, is a reflection of the behavior of the four quantities introduced in equation (\ref{x_op}), which are individually discontinuous at the transition (we provide evidence of this fact in \ref{sec:sm_cons}). In spite of the non-trivial behaviour of $X_C$, the utility of consumers increases monotonically with $n$, in agreement with expectations based on Welfare theorems. Indeed, it is not only the level of consumption that matters, but also its variety.

\begin{figure}
\centering
\includegraphics[width=0.49\columnwidth]{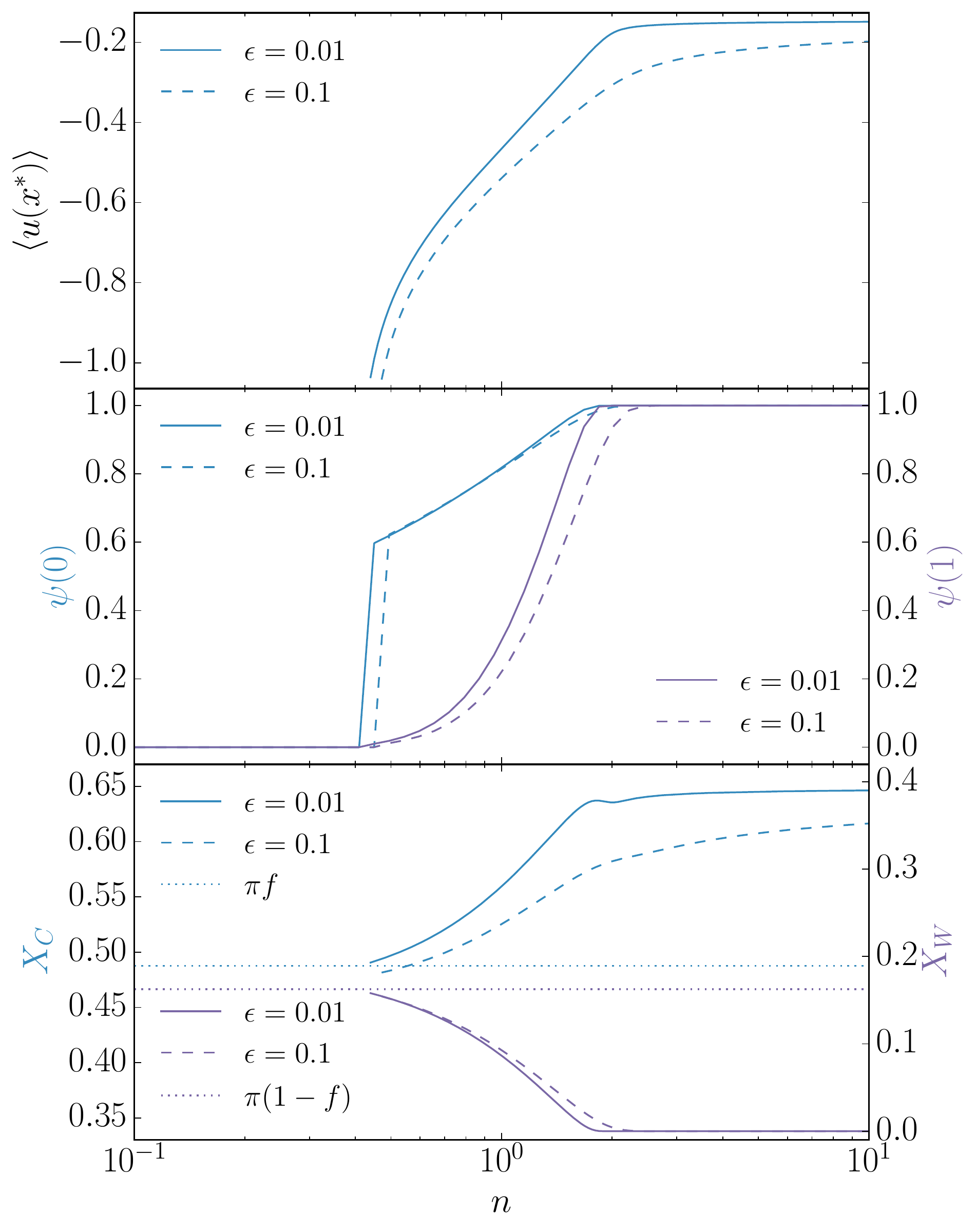}
\caption{\textbf{Regimes of the economy: consumption.} Utility function per final good $\langle u(x^*) \rangle$ (top panel), fraction of efficiently processed intermediate goods $\psi(0)$ and primary non-final goods $\psi(1)$ (middle panel), and levels of consumption $X_C$ and waste $X_W$ as a function of level of development of the economy $n$. $\pi =0.65$, $f=0.75$. In the non-operational phase only primary goods are available, and therefore a fraction $\pi f$ (blue dotted line) of all goods are consumed, while a fraction  $\pi (1-f)$ are wasted (violet dotted line). As production starts, $X_C$ jumps to a larger or smaller value, depending on the inefficiency $\epsilon$, and afterwards $X_C \nearrow n$. $X_W \searrow n$, as more and more non-final goods are used by production processes (see middle panel). Both $\psi(0)$ and $\psi(1)$ undergo a discontinuous change while crossing the transition into an operational economy, but the jump is hardly visible for $\psi(1)$.}
\label{fig:wc_psi}
\end{figure}

Notice that, averaging $x^c$ (see equation (\ref{eq:market_clearing})) over all types of goods one finds that $\langle x^* \rangle = \pi - n \epsilon \langle s^* \rangle$. So there is generally a negative relation between the volume of goods $\langle x^* \rangle$ and the average scale of production $\langle s^* \rangle$. The economy reallocates primary goods to production in such a way as to realise the reduction in $\langle x^* \rangle$ by exploiting the goods that are wasted in the non-industrial phase. The aforementioned peak of $\langle s^* \rangle$ around $n=2$, and the fact that it becomes sharper as $\epsilon$ decreases, are the origin of the non-monotone behaviour of $X_C$ seen in figure \ref{fig:wc_psi}.

Clearly, in the non-industrial phase ($\langle s^*\rangle=0$) the only goods in positive amounts are the primary ones, a fraction $f$ of which are also final goods, whereas the rest (a fraction $\pi (1-f)$ of all goods) is waste. As industrial production sets in, a finite fraction of these wasted primary commodities, starts being employed by technologies (see figure \ref{fig:wc_psi}, middle panel, violet lines). At the same time, a finite fraction of intermediate goods are also recruited in the production process (see figure \ref{fig:wc_psi}, middle panel, blue lines). As a consequence, markets where each of these goods are traded at positive prices emerge. As figure \ref{fig:wc_psi} shows, this change is abrupt: the fraction of intermediate and primary non-final goods that are traded in the economy experiences a jump (though the discontinuity is much sharper for intermediate goods).

\begin{figure}
\centering
\includegraphics[width=0.49\columnwidth]{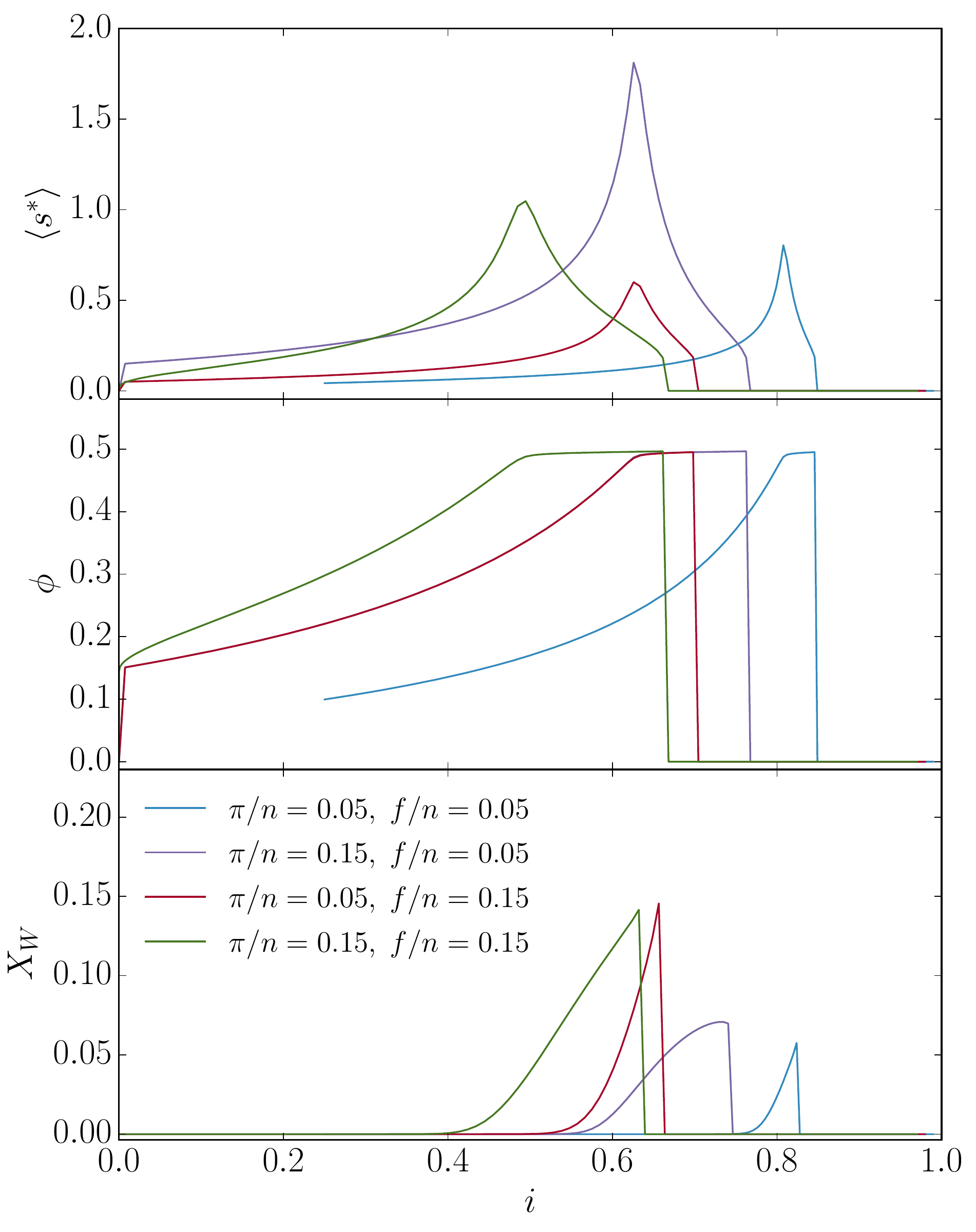}
\caption{\textbf{The role of intermediate goods.} Optimal scales of production $\langle s^* \rangle$ (top panel), fraction $\phi$ of active producers (middle panel), and waste $X_W$ (bottom panel) as a function of the fraction of intermediate goods $i$, for fixed ratios $\pi/n$ and $f/n$ denoting, respectively, the number of primary and final goods over the number of available technologies. $\epsilon=0.1$. We can see that the introduction of primary goods has a beneficial impact on the economy at first, as $\langle s^* \rangle \nearrow i$ and $\phi \nearrow i$. However, as $i$ keeps increasing scales of production peak and the number of active firms saturates. A further increase in $i$ leads to the collapse of the economy, ultimately resulting in $\langle s^* \rangle = 0$ and $\phi = 0$. The increase in $X_W$ signals that non-final goods (including intermediate goods) are not efficiently processed. Such behaviour is consistent with figure \ref{fig:s_phi}, as increasing $i$ at fixed $\pi/n$ amounts to moving along diagonals from the upper right corner to to the bottom left corner of the plane $(n, \pi)$.}
\label{fig:IG}
\end{figure}

While final and primary goods are related to intrinsic properties of the economy (preferences and endowments), one could imagine a scenario in which economic expansion is driven by the proliferation of intermediate goods. These correspond, for example, to services (e.g.\ finance, legal services, etc) or goods produced at intermediate steps in a production chain. It is instructive to analyse the behaviour of the economy as a function of the number of intermediate goods.
This entails looking at the behaviour of the economy at fixed $f/n$ and $\pi/n$, i.e.\ the ratios of the number of primary and final goods to the number of technologies, while the fraction $i=(1-f)(1-\pi)$ of intermediate goods, varies. Figure \ref{fig:IG} shows that the expansion of intermediate goods initially goes along with the expansion in the scales of production (top panel) and with an increase in the number of technologies used (middle panel). For higher value of $i$, $\langle s^*\rangle$ reaches a peak and it decreases for larger $i$. Correspondingly the fraction of operating firms saturates to a constant level. As the number of intermediate goods increases even further, the economy collapses. The reason for this behaviour can be understood by observing that an expansion of the economy through the proliferation of intermediate goods at fixed $f/n$ and $\pi/n$ is achieved by increasing the overall number $C$ of goods in the economy, which in turn causes $n=N/C$ to decrease. In addition, the proliferation of intermediate goods also causes the fraction of primary goods to shrink (indeed, one has $\pi = (1-f-i)/(1-f)$): as shown in figure \ref{fig:s_phi}, both these factors lead the economy towards the shutdown of all production processes.


From the welfare point of view, it is worth to point out that the average utility of consumers {\em decreases} when new goods are introduced, because new constraints are added to the scales of production. Yet, the decrease in $\langle u(x^*)\rangle$ is almost negligible before the peak, while it becomes sharper and sharper as the transition is approached.
The amount of waste exhibits a similar behaviour as the number of intermediate goods increases (figure \ref{fig:IG} bottom panel): For small values of $i$ the waste $X_W$ is almost negligible. On the contrary, beyond the point where $\langle s^*\rangle$ attains its maximum, the waste starts increasing considerably signalling that the economy is not able to process and take advantage of all the intermediate outputs of the production process.


\begin{figure}
\centering
\includegraphics[width=\columnwidth]{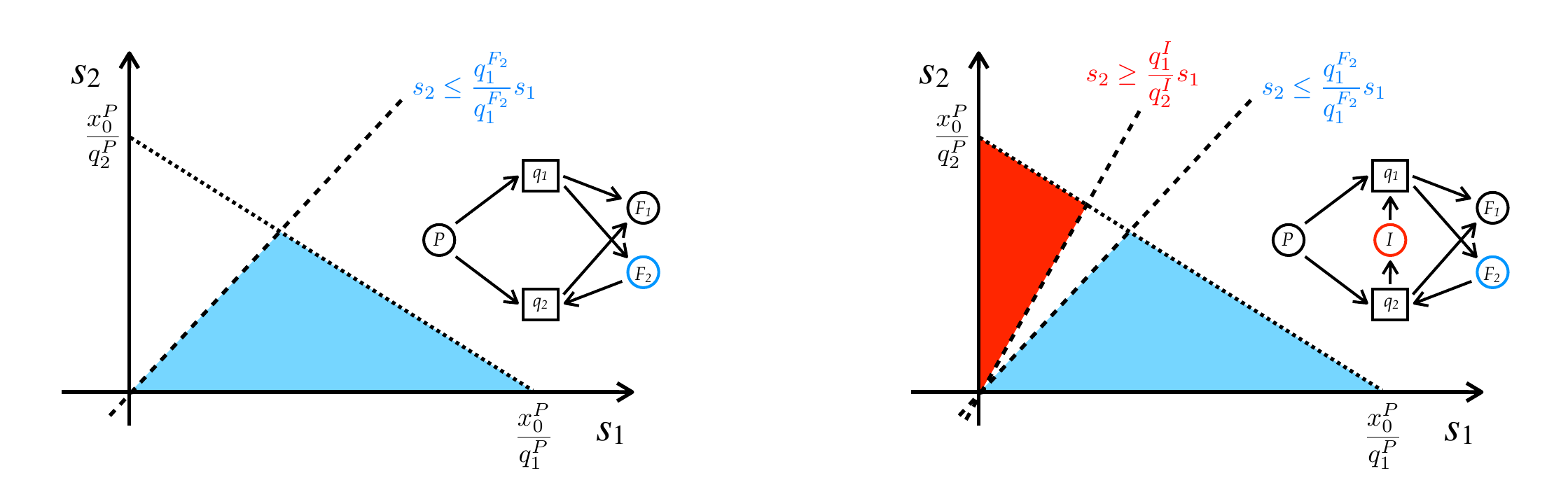}
\caption{\textbf{Geometric interpretation.} Feasible production set for an economy with $N = 2$ technologies, one primary good and two final goods (left panel). The constraint $x^P=x_0^P-q_1^Ps_1-q_2^Ps_2\ge 0$ is satisfied in the region below the dotted line. The good $F_1$ is produced in non-negative amounts for all $s_1,s_2\ge 0$, whereas the constraint $x^{F_2}\ge 0$ singles out the shaded blue region. The introduction of an intermediate good (right panel) introduces one further constraint $x^I\ge 0$ (that is satisfied in the red shaded region). This may cause the collapse of production because the only point where all constraints are satisfied is the origin $s_1=s_2=0$.
Note that primary goods introduce non-homogeneous constraints whereas non-primary goods correspond to homogeneous constraints.}
\label{fig:pictorial}
\end{figure}

\section{Geometric perspective} \label{sec:geometry}
The transition between the industrial and pre-industrial phases has its origin in the constraints $x^c\ge 0~\forall c$. Each of these, in view of equation (\ref{eq:market_clearing}), identifies a hyperplane cutting the $N$-dimensional space of technologies $\boldsymbol{s}$ into a feasible and an unfeasible half-space. The volume $V$ of feasible production scales $\boldsymbol{s}$ corresponds to the intersection between the feasible half-spaces corresponding to all goods $c$ and the positive orthant $\boldsymbol{s}^*\ge 0$.

This construction is sketched in figure \ref{fig:pictorial} (left panel) for a simple economy with one primary good, two final goods and two technologies. Notice that constraints associated with primary goods ($x^c_0 = 1$, dotted line) are \emph{non-homogeneous}, in the sense that they correspond to hyperplanes that do not contain the origin $\boldsymbol{s} = 0$. Constraints associated with non-primary goods ($x^c_0 = 0$, dashed line) are instead \emph{homogeneous} and the corresponding hyperplane contains the origin. Therefore, non-homogeneous constraints can contribute to shrink the volume $V$ but they cannot make it vanish. Conversely, homogeneous constraints select ``slices'' of space, whose intersection can be limited to the origin.
This is shown in figure \ref{fig:pictorial} (right panel) for the case of the simple economy discussed above, when a new intermediate good is introduced. 

Let us consider initially the case in which the signs of $q_1^I$ and $q_2^I$ (and the corresponding arrows in the diagram) are reversed with respect to figure \ref{fig:pictorial} (right panel). In this case the new constraint $x^I \geq 0$ is redundant. In fact, the region of the plane $(s_1, s_2)$ compatible with all the constraints does not change after the introduction of the new intermediate good and the new equilibrium has the same production scales $\boldsymbol{s}^*$ as the prevailing one. In such region $x^I \geq 0$, meaning that in the new equilibrium the new intermediate good is in excess supply and therefore it contributes to waste and has $p^I = 0$. In the case in which the new constraint $x^I \geq 0$ is not redundant, several scenarios can occur instead. The production scales $\boldsymbol{s}^*$ corresponding to the prevailing equilibrium could still be compatible with the new constraint. This case is analogous to the previous one, in the sense that in new equilibrium the scales of production $\boldsymbol{s}^*$ are unchanged and $p^I = 0$. However, if $\boldsymbol{s}^*$ corresponding to the prevailing equilibrium is in the region $x^I < 0$, it means that (from the perspective of the prevailing equilibrium) the new intermediate good is in negative excess supply, i.e.\ in excess demand. The old equilibrium will be displaced and $\boldsymbol{s}^*$ will adjust so that $x^I=0$ and $p^I>0$. However, if the region of the plane $(s_1, s_2)$ compatible with all the constraints reduces to the origin, the economy will collapse to the $\boldsymbol{s}^*=0$ state, as in figure \ref{fig:pictorial} (right panel).

The same intuition carries over to the case of large random economies where the volume $V$ of feasible production vectors $\boldsymbol{s}^*\ge 0$ depends on the particular realization of the random technologies $\boldsymbol{q}$ and endowments $\boldsymbol{x}_0$.
As the number of non-primary goods increases, the set $V$ of feasible production plans shrinks because additional homogeneous constraints are introduced. If the economy has too many non-primary goods (or too few technologies) the volume $V$ ultimately collapses to the single point $\boldsymbol{s} = 0$.

Since the transition depends only on the properties of the production sector, the position of the critical line is independent of the fraction of final goods. The critical line can be computed analytically using the same techniques outlined in section 
\ref{sec:random},
and applied in \cite{BardosciaLivanMarsili,Marsili} to solve a similar problem. We refer the interested reader to \ref{sec:sm_volume} for a full derivation. The critical lines computed analytically in this way are in perfect agreement with the results in the previous subsection and with numerical simulations.

One interesting aspect of the transition is the discontinuous behaviour of $\langle s^* \rangle$ across the transition, shown in figure \ref{fig:s_phi}. This is entirely consistent with the picture outlined above. Indeed the utility function is monotonically increasing, so $\boldsymbol{s}^*$ is expected to lie on the border of the feasible set, where the non-homogeneous constraints are satisfied as equalities. Adding homogeneous constraints, the feasible set becomes a thinner and thinner slice until it reduces to the origin. Yet, before that, the length of $\boldsymbol{s}^*$ remains finite, as it is determined by the non-homogeneous constraints. In order to check the soundness of this picture, we sampled vectors $\boldsymbol{s}$ within the feasibility sets of randomly generated instances of economies with large but finite $N$ and $C$. The shape of the feasibility set can be probed by principal component analysis \cite{Jolliffe} of the sampled vectors' correlation matrix. This analysis confirms the presence of privileged directions in the space of vectors $\boldsymbol{s}$ as the economy approaches the phase transition (e.g.\ by decreasing $\pi$ at fixed $n$), as an exceedingly large fraction of the variability between different feasible production scales $\boldsymbol{s}$ is explained by a single principal component (see the \ref{sec:sm_geometry} for details).


\section{Discussion}
The study of large random economies presented here has its own merit as a reference benchmark with respect to which different approaches may be compared.
Its value stands in the transparency of the assumptions made and in the fact that it captures genuine economic complexity. This section is devoted to comparing the behaviour of an economy, as described in the simplified General Equilibrium setting described here, to the
behaviour of real economies.
The aim is to explore the explicatory power of complex and efficient competitive
markets, as captured by the GET framework, and to compare the emergent non-trivial aspects of industrial dynamics with the empirical evidence discussed in economic literature.

Our framework is a static one. Yet, we could think of a world where, at the beginning of every period, Nature endows consumers with a quantity
$\boldsymbol{x}_0$ of primary goods that are transformed into final goods $\boldsymbol{x}$ by the production sector and consumed by the end of the period. What happens from one period
to the next is that a new technology may be invented (i.e.\ $N\to N+1$) or a new good (i.e. $C\to C+1$) may be introduced. In the following we discuss such possibilities, devoting particular attention to the case of intermediate goods. 
The case of the depletion of natural resources can likewise be analysed by studying the behavior of the economy as a function of $\pi$, but it will not be discussed further.

\subsection{Industralization and natural resources}

First notice that, in this extremely simplified setting, the economy features a no-industrialisation trap, even without increasing returns to scale.
Expansion of the technological repertoire or of the number of primary goods is sufficient to escape the trap, without the need to invoke a ``big push'' \cite{BigPush}.
In our simplified world industrial revolutions would occur as sharp transitions
from an economy based on natural resources to one characterised by mass production of final goods. The main determinants
of this transitions are the span of the repertoire of available technologies and the number of primary goods. The transition occurs as any of the two increases. There is a vast literature on the possible determinants
of the industrial revolution and on why this process occurred earlier in some countries than in others (see e.g. \cite{MSV,Allen}).
Our results suggest that typically a country having access to a larger basket of primary resources (e.g.\ because of its
colonial empire) would cross the transition first with respect to a country with the same repertoire of technologies but with a more limited access to primary goods. This contrasts with the {\em curse of natural resources} that observes that economies rich in natural resources tend to grow at a slower pace \cite{Sachs}. This contrast is only apparent. First because the key variable we consider is the variety of the basket of primary goods, not their abundance. Second, Ref.\ \cite{PapyrakisGerlagh} finds that once indirect effects (e.g.\ corruption, trade openness, etc) that are neglected by the GET framework, are taken into account ``resource abundance has a positive direct impact on growth'' \cite{PapyrakisGerlagh}.

\subsection{Industrial dynamics: positive spillovers vs disruptive technology}

Once an economy has entered its industrialisation phase, what type of industrial dynamics would we typically expect?
This depends on the incentives for R\&D activities. We take a simplified picture of industrial dynamics, where R\&D activity generates
a new (randomly drawn) activity $\boldsymbol{q}_{N+1}$ in an already existing equilibrium with a repertoire of $N$ existing activities (R\&D would
also contribute to make existing technologies more efficient, i.e.\ reducing $\epsilon$. Analogous considerations would apply).
If the new technology generates a positive profit at the current prices, it will be adopted and the equilibrium will be displaced.
Otherwise, the technology will remain idle and the equilibrium will not change.
Moving from one equilibrium to the other, each existing technology shall adjust its scale of production. The adjustment will generate
profits if the corresponding scale of activity $s_i$ increases, and it will generate losses if $s_i$ decreases.

In this caricature of industrial dynamics, the regions $n<2$ and $n>2$ markedly differ in terms of the private sector's capabilities to sustain R\&D activities.
Indeed, in the operational phase for $n<2$ new technologies are adopted with high probability ($\phi\simeq 1/2$) and the fact that $\langle s^*\rangle$ increases with $n$ implies that profits will be shared with the rest of the production sector. Such spillovers are endogenous in this phase of the economy. On the contrary, for $n>2$ a new technology will be adopted with a probability that decreases with $n$. In addition, a new successful technology displaces already existing ones (because $\phi\searrow n$) and decreases their average profitability (because $\langle s^*\rangle\searrow n$).
Even though investment is not included in our model, the fact that R\&D activities generate profits in the $n<2$ phase, suggests that they can be sustained in this phase. On the contrary, R\&D activities can hardly be sustained when $n>2$.
It is suggestive to relate this finding to the observation that the rate of introduction of new drugs in a very technologically intensive domain as the pharmaceutical sector, has been constant for decades, in spite of increasing investments in R\&D and of spectacular technological advances \cite{drugdesign}.

\subsection{The role of intermediate goods: vertical integration vs outsourcing}

A second mode of industrial change is the introduction of a new non-final good~\footnote{We mainly discuss the case of intermediate goods. Notice that consumers' preferences do not change when a new non-final good is introduced.}. Examples include outsourcing the production process of intermediate components to external firms with interactions mediated by market prices,
emission permits for pollutants (e.g.\ carbon, nitrogen oxides), services and financial products. In the simplified GET framework discussed here, the sole change is that a new market is created for the new good and each technology acquires a component specifying its contribution to the production or usage of that good. For a new primary good, the endowment vector $\boldsymbol{x}_0$ acquires an additional non-zero component.
As discussed earlier, if the new good is in excess supply the prevailing equilibrium is not modified, whereas if the new good is in negative excess supply~\footnote{We stress again that the new good can be in negative excess supply only with respect to the prevailing equilibrium, which will then be displaced in favour of a new equilibrium in which the new good will be either in positive excess supply or fully utilized by production processes.} (e.g.\ carbon emissions) then the economy's equilibrium will adjust so as to clear the market.

Again, the regions $n<2$ and $n>2$ markedly differ in terms of the incentives they generate to support this process. For $n<2$ adding a new good comes with a reduction of the average scale $\langle s^*\rangle$ of operations, because the increase in the number of (intermediate) goods implies a reduction in $n$. This implies losses for the productive sector. This suggests that {\em vertically integrated} production processes should prevail in the early stages of industrialisation ($n<2$). For an economy in an advanced stage of industrialisation ($n>2$) instead, the incentives for outsourcing are positive, again because $\langle s^*\rangle \searrow n$. This is remarkably reminiscent of the
 account that Langlois \cite{Langlois} offers of industrial evolution in the last four centuries. In brief, industrial dynamics has been dominated for long by vertically integrated firms, that incorporated all stages of production. This required intensive managerial skills to buffer the volatility inherent in the intermediate stages of the production process. As demand expanded, market institutions developed in order to support stable prices. The resulting decrease in volatility made it profitable to outsource part of the production process by generating new competitive markets for intermediate goods. The same argument suggests that the introduction of carbon emission trading systems may generate profits in the industrial sectors of advanced economies ($n>2$) and losses in those of developing ones ($n<2$).

From the welfare point of view, it has to be observed that while technological innovation (i.e.\ increasing $N$) always leads to an improvement in consumers' welfare, the opposite is true for the introduction of new intermediate goods (i.e.\ increasing $C$), because it imposes further constraints on the set of feasible production scales $\boldsymbol{s}$. These welfare changes are substantial for $n<2$ whereas they are smaller for $n>2$.

In summary, the statistical mechanics approach to the GET of large random economies exhibits a rich behaviour that informs us on
what competitive markets can achieve in typical cases, without invoking
non-equilibrium effects, equilibrium multiplicity (e.g.\ increasing returns) or market inefficiencies.
It ultimately suggests that the statistical mechanics of large random economies possesses an explicative potential that is yet untapped.
\\

\section*{Acknowledgments}
All authors would like to thank Federico Ricci-Tersenghi and Davide Fiaschi for stimulating discussions, and would like to acknowledge support from the FoodCAST project (SISSA). Marco Bardoscia acknowledges support from: FET Project SIMPOL nr.\ 610704, FET project DOLFINS nr.\ 640772. Giacomo Livan acknowledges support from an EPSRC Early Career Fellowship (Grant No.\ EP/N006062/1).


\appendix

\section{The optimization problem} \label{sec:sm_optm}
In this section we show how the typical properties of the economy (i.e.\ properties attained by each random realization of the economy in the limit $N,C \rightarrow \infty$ with $n = N/C$ fixed) can be computed. We start from the maximization of the utility function (equation (\ref{eq:max})). 
For later convenience, we write this as:
\begin{equation} \label{utility_2}
U(\boldsymbol{s}) = \sum_{c = 1}^C k^c \ u \left( x_0^c + \sum_{i=1}^N q_i^c s_i \right) \,
\end{equation}
where $k^c$ distinguishes final and non-final goods, i.e.\ $k^c = 1$ if $c \in \mathcal{F}$, and $k^c = 0$ otherwise. 

We shall compute the typical properties under the following assumptions on the economy's parameters:
\begin{itemize}
\item $x_0^c$ denotes good $c$'s initial endowment. Each good is primary (and hence part of the initial endowments) with probability $\pi$ and independently from the other goods, i.e.\ initial endowments follow a bimodal probability density distribution
\begin{equation} \label{primary_prob}
\rho_\mathcal{P} (x_0^c) = (1-\pi) \delta(x_0^c) + \pi \delta(x_0^c-1) \ ,
\end{equation}
where $\delta (\cdot)$ denotes Dirac's delta.
\item Each good is final (and hence part of the utility function) with probability $f$ and independently from the other goods, i.e.\ the auxiliary variables $k^c$ introduced above follow the bimodal density
\begin{equation} \label{final_prob}
\rho_\mathcal{F}(k^c) = (1-f) \delta(k^c) + f \delta(k^c-1) \ .
\end{equation}
\item The economy's input--output matrix $\boldsymbol{q} = \{ q_i^c \}_{i=1,\ldots,N}^{c=1,\ldots,C}$ is defined as a set of constrained Gaussian random numbers with mean zero and variance $1/C$ such that $\sum_{i=1}^N q_i^c = -\epsilon$, $\forall \ c$, where $\epsilon > 0$ quantifies the economy's inefficiency.
\end{itemize}

Under the above distributional assumptions, we compute the economy's typical properties by resorting to techniques borrowed from the statistical mechanics of disordered systems. Namely, let us consider a utility function $U_N(\boldsymbol{s} | \boldsymbol{x}_0, \boldsymbol{q}, \boldsymbol{k})$ for a system of given size, i.e.\ with $N$ technologies and $C$ goods ($n = N/C$) and for a given realization of the input--output matrix $\boldsymbol{q}$, the initial endowments $\boldsymbol{x}_0 = (x_0^1, \ldots, x_0^C)$, and the consumer good labels $\boldsymbol{k} = (k^1, \ldots, k^C)$. Given the additivity of the utility function in equation (\ref{utility_2}), one expects its maxima to grow with $N$. Thus, we shall attempt at solving the following maximization problem:
\begin{equation} \label{max1}
\lim_{N \rightarrow \infty} \frac{1}{N} \max_{\boldsymbol{s} \geq 0} U_N (\boldsymbol{s} | \boldsymbol{q}, \boldsymbol{x}_0, \boldsymbol{k}) = \lim_{N \rightarrow \infty} \lim_{\beta \rightarrow \infty} \frac{1}{\beta N} \log Z_N (\beta | \boldsymbol{q}, \boldsymbol{x}_0, \boldsymbol{k}) \ ,
\end{equation} 
where
\begin{equation} \label{partfunc}
Z_N (\beta | \boldsymbol{q}, \boldsymbol{x}_0, \boldsymbol{k}) = \int \mathrm{d} \boldsymbol{s} \ \mathrm{e}^{\beta U_N (\boldsymbol{s} | \boldsymbol{q}, \boldsymbol{x}_0, \boldsymbol{k})} \ ,
\end{equation}
and $\mathrm{d} \boldsymbol{s} = \prod_{i=1}^N \mathrm{d}s_i$. Thus, the optimization problem in equation (\ref{max1}) amounts to converting the maximization problem into a steepest descent problem on the integral in equation (\ref{partfunc}): the limit $\beta \rightarrow \infty$ selects regions in the $\mathbf{s}$ space where $U_N$ takes its largest values. Also, for a smooth enough $U_N$, one can safely expect that, when $N$ becomes large, the solutions of the maximization problem do not depend on the specific realization of random variables, i.e.\ one can expect $\max_{\boldsymbol{s}} U_N /N$ to become a self-averaging quantity:
\begin{equation} \label{max2}
\lim_{N \rightarrow \infty} \frac{1}{N} \max_{s} U_N (\boldsymbol{s} | \boldsymbol{q}, \boldsymbol{x}_0, \boldsymbol{k}) = \lim_{N \rightarrow \infty} \frac{1}{N} \left \langle \max_{s} U_N (\boldsymbol{s} | \boldsymbol{q}, \boldsymbol{x}_0, \boldsymbol{k}) \right \rangle_{\boldsymbol{q}, \boldsymbol{x}_0, \boldsymbol{k}} \ .
\end{equation}
The averaging operation in the above equation makes it difficult to exploit the identity in equation (\ref{max1}), as computing the average of the logarithm of the partition function $Z_N$ is typically a very difficult task. This problem can be circumvented by resorting to the \emph{replica trick}, i.e.\ by exploiting the following identity (where we omit the conditioning on $\boldsymbol{q}, \boldsymbol{x}_0, \boldsymbol{k}$) one can convert the averaging over the partition function logarithm into an averaging over the partition function of $r$ replicas of the same system:
\begin{equation} \label{replica_trick_1}
\left \langle \log Z_N \right \rangle = 
\lim_{r \rightarrow 0} \frac{\langle Z_N^r \rangle-1}{r}  \ .
\end{equation}
In our case, the partition function in equation (\ref{partfunc}) reads
\begin{equation} \label{partfunc2}
Z_N (\beta | \boldsymbol{q}, \boldsymbol{x}_0, \boldsymbol{k}) = \int_0^\infty \mathrm{d} \boldsymbol{s} \ \exp \left \{ \beta \sum_{c=1}^C k^c u \left( x_0^c + \sum_{i=1}^N q_i^c s_i \right)  \right \} \ .
\end{equation}

By closely following the derivation in \cite{DeMartinoMarsiliPeresCastillo}~\footnote{The utility function in \cite{DeMartinoMarsiliPeresCastillo} is the same used in this paper with $k^c = 1$ $\forall c$. Hence, in our case it is sufficient to perform the two replacements $u(x^c) \to k^c u(x^c)$ and $\langle \ldots \rangle_{t,x_0} \to \langle \ldots \rangle_{t,x_0,k}$.}, one eventually gets to the following result
\begin{equation} \label{solution}
\lim_{N \rightarrow \infty} \frac{1}{N} \left \langle \max_{s} U_N (\boldsymbol{s} | \boldsymbol{q}, \boldsymbol{x}_0, \boldsymbol{k}) \right \rangle_{\boldsymbol{q}, \boldsymbol{x}_0, \boldsymbol{k}} = h(\Omega,\kappa,p,\sigma,\chi,\hat{\chi})
\end{equation}
where $\Omega,\kappa,p,\sigma,\chi,\hat{\chi}$, are self-consistently determined by setting the partial derivatives of $h$ with respect to such parameters equal to zero. In the jargon of statistical mechanics these variables are known as order parameters, while the six resulting equations are called saddle point equations. The function $h$ reads:
\begin{eqnarray} \label{heq}
h(\Omega,\kappa,p,\sigma,\chi,\hat{\chi}) &=& \left \langle \max_{s \geq 0} \left [ - \frac{\hat{\chi}}{2} s^2 + (t\sigma - p \epsilon) s \right ] \right \rangle_t + \frac{1}{2} \Omega \hat{\chi}  \\ \nonumber
&-& \frac{1}{2} \frac{\chi}{n\Delta} \left (\sigma^2 + p^2 \right ) + \frac{1}{n} \kappa p \\ \nonumber &+& \frac{1}{n} \left \langle \max_{x \geq 0} \left [ k u(x) - \frac{(x - x_0 + \kappa + \sqrt{n\Omega} t)^2}{2\chi} \right ] \right \rangle_{t,x_0,k},
\end{eqnarray}
where $t$ is a standard Gaussian random variable, and $\langle \ldots \rangle_{t,x_0,k}$ denotes the average over such variable, initial endowments, and labels $k$ distributed according to equations (\ref{primary_prob}) and (\ref{final_prob}), respectively. The above equation can be written as
\begin{eqnarray} \label{heq2}
h(\Omega,\kappa,p,\sigma,\chi,\hat{\chi}) &=& - \frac{\hat{\chi}}{2} \left \langle (s^*)^2 \right \rangle_t + \sigma \left \langle t s^* \right \rangle_t - p \epsilon \langle s^* \rangle_t + \frac{1}{2} \Omega \hat{\chi} \\ \nonumber &-& \frac{1}{2} \frac{\chi}{n} \left (\sigma^2 +  p^2 \right )
+ \frac{1}{n} \kappa p + \frac{1}{n} \langle k u(x^*) \rangle_{t,x_0,k} \\ \nonumber &-& \frac{1}{2n\chi} \left \langle (x^* - x_0 + \kappa + \sqrt{n \Omega} t)^2 \right \rangle_{t,x_0,k} \ ,
\end{eqnarray}
where $s^*$ and $x^*$ denote the solutions of the two maximization problems in equation (\ref{heq}). $s^*$ is found by setting the derivative of the first term in equation (\ref{heq}) equal to zero:
\begin{equation} \label{sstar}
s^* = \frac{\sigma t - p \epsilon}{\hat{\chi}} \, \Theta \left ( \sigma t - p \epsilon \right ) \ ,
\end{equation}
where the Heaviside $\Theta$ function guarantees that $s^*$ is non-negative. In fact, as noted in the main text, if a technology becomes unprofitable, it is shut down by posing the corresponding scale of production equal to zero. On the other hand, $x^*$ needs to be computed as a solution to the following equation:
\begin{equation} \label{xstar}
k \chi u^\prime(x^*) = x^* - x_0 + \kappa + \sqrt{n \Omega}t \ ,
\end{equation}
with the constraint that $x^*$ must be positive. According to the above relation, $x^*$ can take the following values:
\begin{equation} \label{xstarcases}
x^* = \left \{
\begin{array}{ll}
	\chi u^\prime(x^*) + x_0 -\kappa - \sqrt{n\Delta \Omega} t & \qquad \mathrm{for} \ k = 1 \\
	(x_0 - \kappa - \sqrt{n\Delta \Omega}t) \Theta(x_0 - \kappa - \sqrt{n\Delta \Omega}t) & \qquad \mathrm{for} \ k = 0 \ .
\end{array} \right.
\end{equation}
The positivity of $x^*$ must be explicitly enforced by the Heaviside $\Theta$ function for $k=0$, whereas for $k=1$ the presence of an explicit constraint might be redundant. In particular, if $\chi > 0$, the standard choice $u(x) = \log(x)$ leads to a quadratic equation for $x^*$, that always has the following positive solution:
\begin{equation} \label{xstar}
x^* = \frac{1}{2} \left [ x_0 - \kappa - \sqrt{n \Omega} t + \sqrt{(x_0 - \kappa - \sqrt{n \Omega} t)^2 + 4 \chi} \right ] \ .
\end{equation}
However, in the case $\chi = 0$ there is no difference between the two cases $k = 0$ and $k = 1$, and the solution simply is $x^* = (x_0 - \kappa - \sqrt{n \Omega}t) \Theta(x_0 - \kappa - \sqrt{n \Omega}t)$. With these positions, the saddle point equations read:
\begin{eqnarray} \label{sp}
p &=& \frac{1}{\chi} \mathcal{M}_1(x^*,\kappa,\Omega) \\ \nonumber
\hat{\chi} &=& \frac{1}{\sqrt{{n\Omega\chi^2}}} \mathcal{M}_t(x^*,\kappa,\Omega) \\ \nonumber
\sigma &=& \sqrt{\frac{\mathcal{M}_2(x^*,\kappa,\Omega)}{\chi^2} - p^2} \\ \nonumber
\Omega &=& \langle (s^*)^2 \rangle_t \\ \nonumber
\kappa &=& p\chi + n\epsilon \langle s^* \rangle_t \\ \nonumber
\chi &=& \frac{n}{\sigma} \langle s^* t \rangle_t \ ,
\end{eqnarray}
where
\begin{eqnarray} \label{avg}
\mathcal{M}_1(x^*,\kappa,\Omega) &=& \left \langle x^* - x_0 + \kappa + \sqrt{n \Omega}t \right \rangle_{t,x_0,k} \\ \nonumber
\mathcal{M}_t(x^*,\kappa,\Omega) &=& \left \langle \left (x^* - x_0 + \kappa + \sqrt{n \Omega}t \right ) t \right \rangle_{t,x_0,k} \\ \nonumber
\mathcal{M}_2(x^*,\kappa,\Omega) &=& \left \langle \left (x^* - x_0 + \kappa + \sqrt{n \Omega}t \right )^2 \right \rangle_{t,x_0,k} \ ,
\end{eqnarray}
and where the average over $k$ is non-trivial because $x^*$, the solution of equation (\ref{xstarcases}) depends on $k$. The simultaneous solution of the above equations yields the values of the order parameters, which in turn can be used to compute the model's quantities of interest (such as $\langle s^* \rangle$ and $\langle x^* \rangle$), as discussed in the following. It is worth highlighting that the original problem in equation (\ref{max1}), which entails optimization over an infinite number of variables (since $N$ approaches infinity), has been reduced to a set of six nonlinear equations.  

Let us focus on $\mathcal{M}_1$ to see how to compute the above quantities:
\begin{eqnarray} \label{M1}
\mathcal{M}_1(x^*,\kappa,\Omega) &=& f \left \langle x^* - x_0 + \kappa + \sqrt{n \Omega} t \right \rangle_{t,x_0,k=1} + \\ \nonumber
& & + (1-f) \left \langle x^* - x_0 + \kappa + \sqrt{n \Omega} t \right \rangle_{t,x_0, k=0} \\ \nonumber
&=& f \chi \langle u^\prime(x^*) \rangle_{t,x_0} \\ \nonumber
& & + (1-f) \left \langle (\kappa + \sqrt{n \Omega} t - x_0) \Theta (\kappa + \sqrt{n \Omega} t - x_0) \right \rangle_{t,x_0}.
\end{eqnarray}
Following this line of reasoning, and explicitly evaluating, when possible, the averages over $t$ we get to:
\begin{eqnarray} \label{Meq}
\mathcal{M}_1(x^*,\kappa,\Omega) &=& f \chi \langle u^\prime(x^*) \rangle_{t,x_0} + (1-f) \langle \mathcal{I}_1 (x_0,\kappa,\Omega) \rangle_{x_0} \\ \nonumber
\mathcal{M}_t(x^*,\kappa,\Omega) &=& f \chi \langle u^\prime(x^*) t \rangle_{t,x_0} + (1-f) \langle \mathcal{I}_t (x_0,\kappa,\Omega) \rangle_{x_0} \\ \nonumber
\mathcal{M}_2(x^*,\kappa,\Omega) &=& f \chi^2 \left \langle (u^\prime(x^*))^2 \right \rangle_{t,x_0} + (1-f) \langle \mathcal{I}_2 (x_0,\kappa,\Omega) \rangle_{x_0} \ ,
\end{eqnarray}
where
\begin{eqnarray} \label{Ieq}
\mathcal{I}_1(x_0,\kappa,\Omega) &=& \sqrt{\frac{n \Omega}{2\pi}} \exp \left ( -\frac{(x_0-\kappa)^2}{2n \Omega} \right ) -
 (x_0-\kappa) \ \psi(x_0, \kappa, \Omega) \\ \nonumber
\mathcal{I}_t(x_0,\kappa,\Omega) &=& \sqrt{n \Omega} \ \psi(x_0, \kappa, \Omega) \\ \nonumber
\mathcal{I}_2(x_0,\kappa,\Omega) &=& n \Omega \left [ \left ( 1 + \frac{(x_0-\kappa)^2}{n \Omega} \right ) \psi(x_0, \kappa, \Omega) \right. \\ \nonumber
& & \qquad \left. - \frac{x_0-\kappa}{\sqrt{2\pi n \Omega}} \exp \left ( - \frac{(x_0-\kappa)^2}{2n \Omega} \right ) \right ] \ 
\end{eqnarray}
with
\begin{equation} \label{eq:psi}
\psi(x_0, \kappa, \Omega) = \frac{1}{2} \mathrm{erfc} \left( \frac{x_0 - \kappa}{\sqrt{2 n \Omega}}\right) \ ,
\end{equation}
which, as we will show in the following (see equation (\ref{xstar_dist_k1})), is simply the fraction of efficiently processed intermediate goods. \\

As already mentioned, any quantity of interest is a function of the order parameters, which, in turn, must be computed by solving the saddle point equations (\ref{sp}). Solutions to such equations where all order parameters attain finite values are found only in a certain region of the $(n,\pi)$ plane. This is a symptom of the phase transition we present in the paper, and which we fully characterize analytically in \ref{sec:sm_volume} and \ref{sec:sm_geometry}. At this level, the emergence of the transition can be linked to the behavior of the order parameter $\chi$. In fact, as shown in \cite{Marsili} we have
\begin{equation} \label{chi}
\chi = \frac{\beta n}{2N} \sum_{i=1}^N (s_{i,a} - s_{i,b})^2 \ ,
\end{equation}
where the indices $a$ and $b$ denote two different replicas~\footnote{In order to proceed from (\ref{replica_trick_1}) to (\ref{solution}), one has to perform a replica symmetric ansatz, which amounts to assuming that the distances between different pairs of replicas are equal.}. The above quantity is affected by two ``competing'' limits, as both $\beta$ and $N$ grow to infinity. The limit for $\beta \to \infty$ selects the solution of the optimization problem in equation (\ref{max1}), while the limit $N \to \infty$ ensures that such a solution is typical. Hence, under such limits different replicas converge to the same solution, so that the average distance between replicas $\sum_{i=1}^N (s_{i,a} - s_{i,b})^2 / N$ becomes vanishingly small. In particular, in order for $\chi$ to attain a finite value the average distance between replicas must decay as $\beta^{-1}$ for large $\beta$. As we shall show, the phase transition presented in the paper takes place when the only acceptable solution to the optimization problem is $\boldsymbol{s} = 0$, which implies $\chi = 0$. \\

For $\chi = 0$ one has to recompute the averages in equation (\ref{avg}). It is easy to show that in this case equation (\ref{Meq}) greatly simplifies:
\begin{eqnarray} \label{Meq2}
\mathcal{M}_1(\kappa,\Omega) &=& \langle \mathcal{I}_1 (x_0,\kappa,\Omega) \rangle_{x_0} \\ \nonumber
\mathcal{M}_t(\kappa,\Omega) &=& \langle \mathcal{I}_t (x_0,\kappa,\Omega) \rangle_{x_0} \\ \nonumber
\mathcal{M}_2(\kappa,\Omega) &=& \langle \mathcal{I}_2 (x_0,\kappa,\Omega) \rangle_{x_0} \ .
\end{eqnarray}
Introducing the rescaled parameters
\begin{equation} \label{rescalechi}
\ell = p \chi \ , \qquad \gamma = \sigma \chi \ , \qquad \delta = \hat{\chi} \chi \ ,
\end{equation}
the saddle point equations can be rewritten for $\chi = 0$ as follows:
\begin{eqnarray} \label{sp_chizero}
\ell &=& \mathcal{M}_1(\kappa,\Omega) \\ \nonumber
\delta &=& \frac{1}{\sqrt{{n\Omega}}} \mathcal{M}_t(\kappa,\Omega) \\ \nonumber
\gamma &=& \sqrt{\mathcal{M}_2(\kappa,\Omega) - \ell^2} \\ \nonumber
\Omega &=& \langle (s^*)^2 \rangle_t \\ \nonumber
\kappa &=& \ell + n\epsilon \langle s^* \rangle_t \, .
\end{eqnarray}
Let us mention that, in the above list, only $\Omega$ has a straightforward interpretation in terms of fluctuations of the optimal scales of production. It should also be noted that in Ref. \cite{DeMartinoMarsiliPeresCastillo} the parameters corresponding to $p$ and $\sigma$ (see Eqs. (\ref{sp}) and (\ref{rescalechi})) could be interpreted in terms, respectively, of the average and standard deviation of goods' prices. Such an interpretation, however, does not fully translate to the model at hand due to the presence of non-final goods, for which consumer markets and consumer prices are not defined.

As is clear from inspection of equations (\ref{Meq2}) and (\ref{sp_chizero}), the solution of the optimization problem at $\chi = 0$ does not depend on $f$. Hence, the position of the critical line in the plane $(n,\pi)$ does not depend on $f$ as well. Incidentally, we note that it is also possible to show that the above saddle point equations actually reduce to a system of three equations in the variables $\kappa$, $\Omega$ and $\pi$ (the latter comes from the averages over $x_0$) in equation (\ref{Meq2}).

The full distributions of $s^*$ and $x^*$ can be computed from their expressions in equations (\ref{sstar}) and (\ref{xstar}), respectively, by averaging over the standard Gaussian variable $t$. This yields
\begin{equation} \label{sstar_dist}
P(s^*) =  (1-\phi(p, \sigma)) \delta(s^*) + \Theta(s^*) \frac{\hat{\chi}}{\sqrt{2\pi}\sigma} \exp \left( - \frac{(\hat{\chi}s^* + \epsilon p)^2}{2\sigma^2} \right),
\end{equation}
with $p$, $\sigma$ and $\hat{\chi}$ being solutions of equation (\ref{sp}), and $\phi$, which from equation (\ref{sstar_dist}) can be interpreted as the fraction of $s^*$ larger than zero, equal to:
\begin{equation} \label{phi}
\phi(p, \sigma) = \frac{1}{2} \mathrm{erfc} \left( \frac{\epsilon p}{\sqrt{2}\sigma} \right) .
\end{equation}

As regards the distribution of $x^*$ (conditional on $x_0$) one can write:
\begin{eqnarray} \label{xstar_dist}
P(x^* | x_0) &=& P(x^* | x_0, k=1) P(k=1) + P(x^* | x_0, k=0) P(k=0) \\ \nonumber
&=& f P(x^* | x_0, k=1) + (1-f) P(x^* | x_0, k=0),
\end{eqnarray}
where $P(x^* | x_0, k)$ read
\begin{eqnarray} \label{xstar_dist_k1}
P(x^* | x_0, k=1) &=& \frac{1 - \chi u^{\prime\prime}(x^*)}{\sqrt{2\pi n\Omega}} \exp \left( - \frac{(x^* - x_0 - \chi u^\prime(x^*) + \kappa)^2}{2n\Omega} \right) \\ \nonumber
P(x^* | x_0, k=0) &=& \psi(x_0, \kappa, \Omega) \delta(x^*) \\ \nonumber
& & + \Theta(x^*) \frac{1}{\sqrt{2 \pi n \Delta \Omega}} \exp \left( - \frac{(x^* - x_0 +\kappa)^2}{2 n \Delta \Omega} \right)
\end{eqnarray}
with $\kappa$, $\Omega$ and $\chi$ being solutions of equation (\ref{sp}), and $\psi$ has been introduced in equation (\ref{eq:psi}). 

%
The probability densities in equations (\ref{sstar_dist}) and (\ref{xstar_dist}) yield the following expression for the average values
\begin{eqnarray} \label{sstar_avg}
\langle s^* \rangle &=& \frac{\sigma}{\sqrt{2 \pi}\hat{\chi}} \exp \left( - \frac{\epsilon^2 p^2}{2\sigma^2} \right) - \frac{\epsilon p}{\hat{\chi}} \phi(p, \sigma) \\ \nonumber
\langle x^* | x_0 \rangle_{k=0} &=& \sqrt{\frac{n\Delta\Omega}{2 \pi}} \exp \left( - \frac{(x_0 - \kappa)^2}{2n\Delta\Omega} \right) + (x_0 - \kappa)(1 - \psi(x_0, \kappa, \Omega)) \\
\end{eqnarray}
while the average $\langle \ldots | x_0 \rangle_{k=1}$ has to be computed numerically.

In figure \ref{num_fig} we present a numerical check of the solution computed above, which is in excellent agreement with results obtained from a fully numerical maximization of finite sized instances of the problem in equation (\ref{max1}).
\begin{figure}
\centering
\includegraphics[width=0.48\columnwidth]{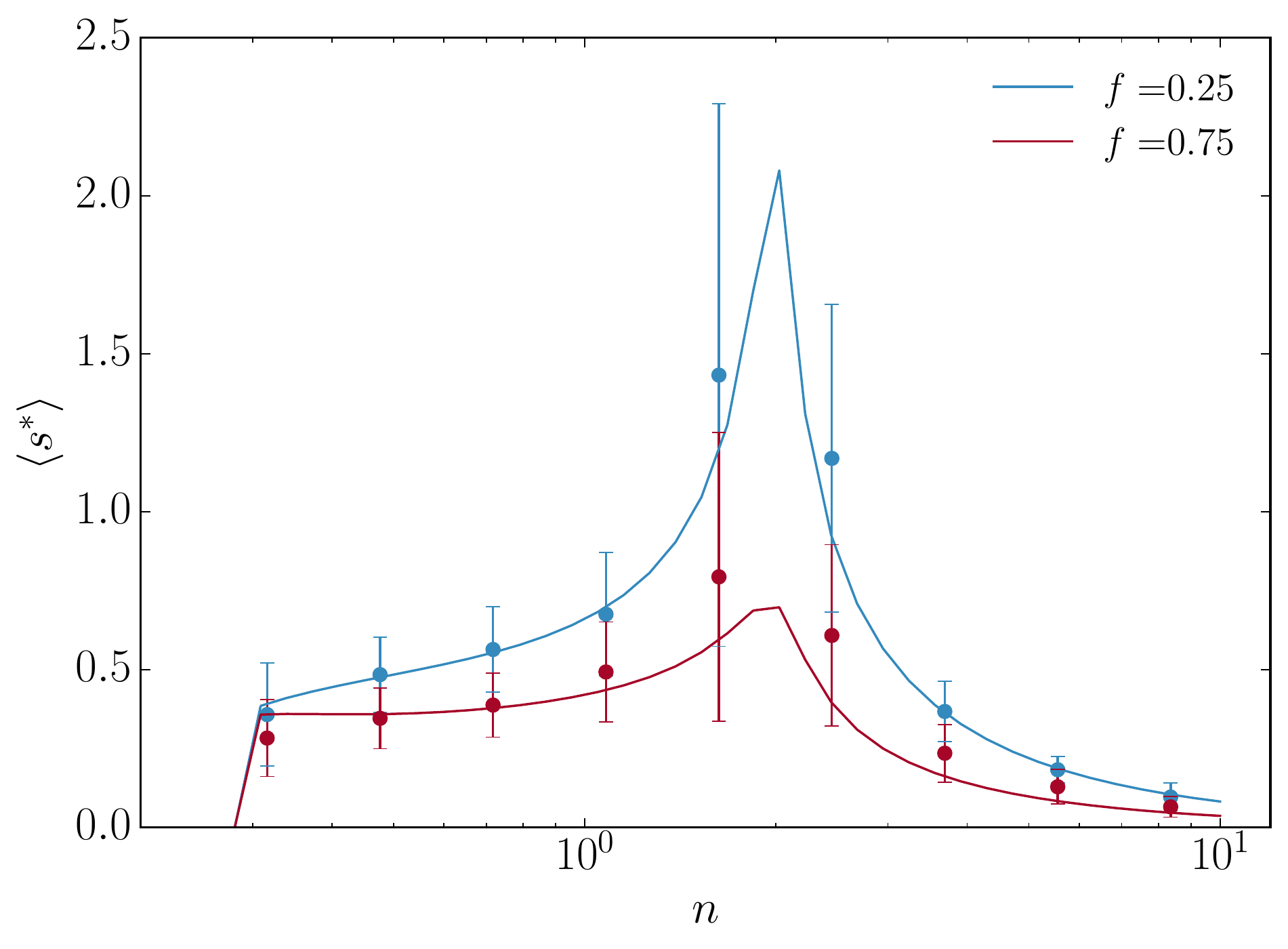}
\includegraphics[width=0.48\columnwidth]{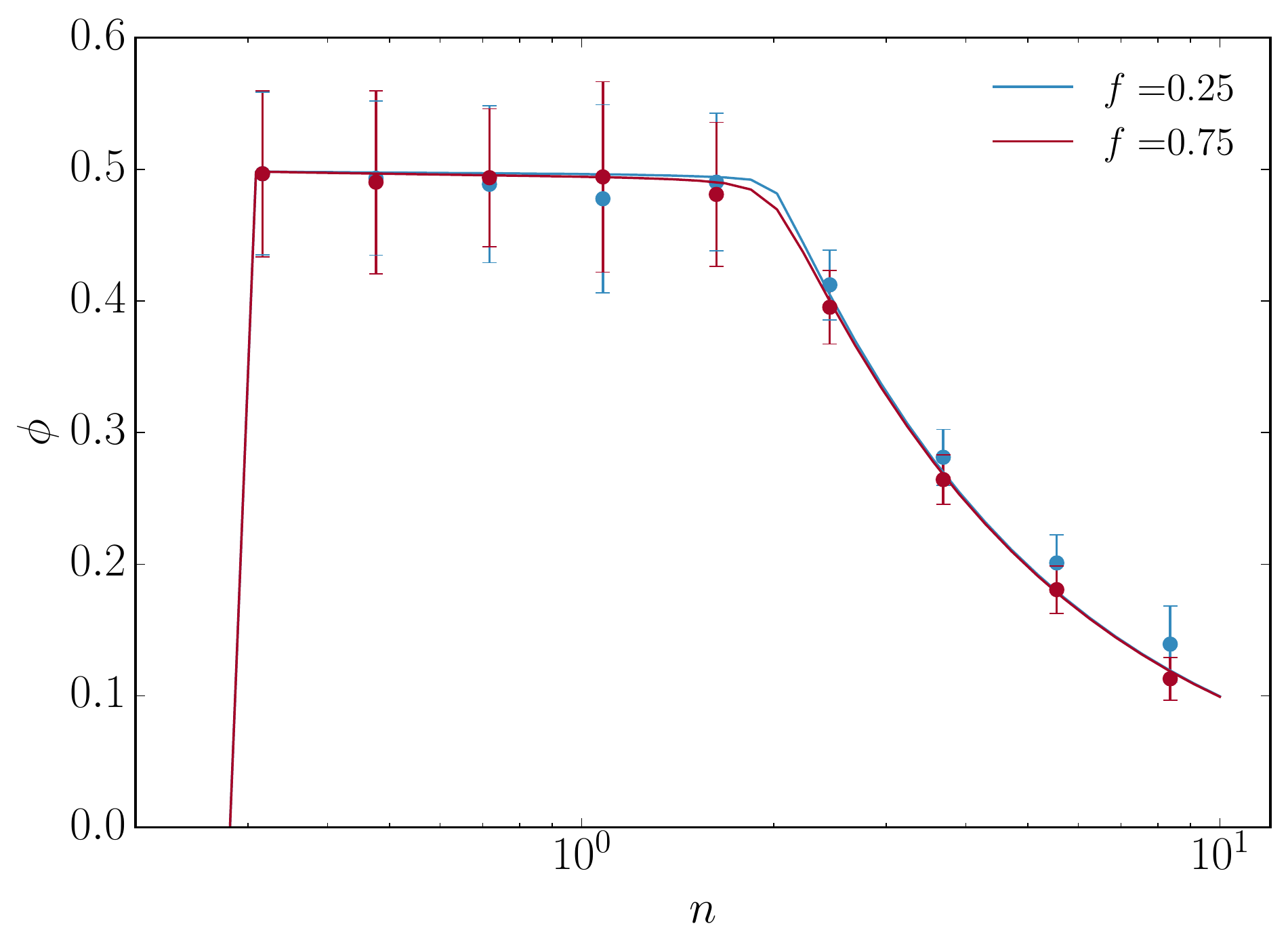}
\caption{\textbf{Numerical verification.} Comparison between $\langle s^* \rangle$ and $\phi$ as computed from equations (\ref{sstar_avg}) and (\ref{phi}) (solid line), respectively, and results from numerical solutions of the optimization problem in equation (\ref{max1}) with finite size $N = 100$ (dots and error bars correspond to mean and standard deviation over a sample of 100 realizations of the random variables $q_i^c$, $x_0^c$ and $k^c$). In both panels $\epsilon = 0.1$, $\pi = 0.65$.}
\label{num_fig}
\end{figure}

As a final remark let us point out the effects of the inefficiency parameter $\epsilon$ on the sharpness of the peak observed at $n \simeq 2$ shown in 
figure \ref{fig:s_phi}, 
which defines the onset of the economy's highly competitive regime. As shown in figure \ref{eps_peak}, the peak sharpens as $\epsilon$ becomes smaller, signaling that for more efficient economies the transition towards competitiveness is abrupt.
\begin{figure}
\centering
\includegraphics[width=0.48\columnwidth]{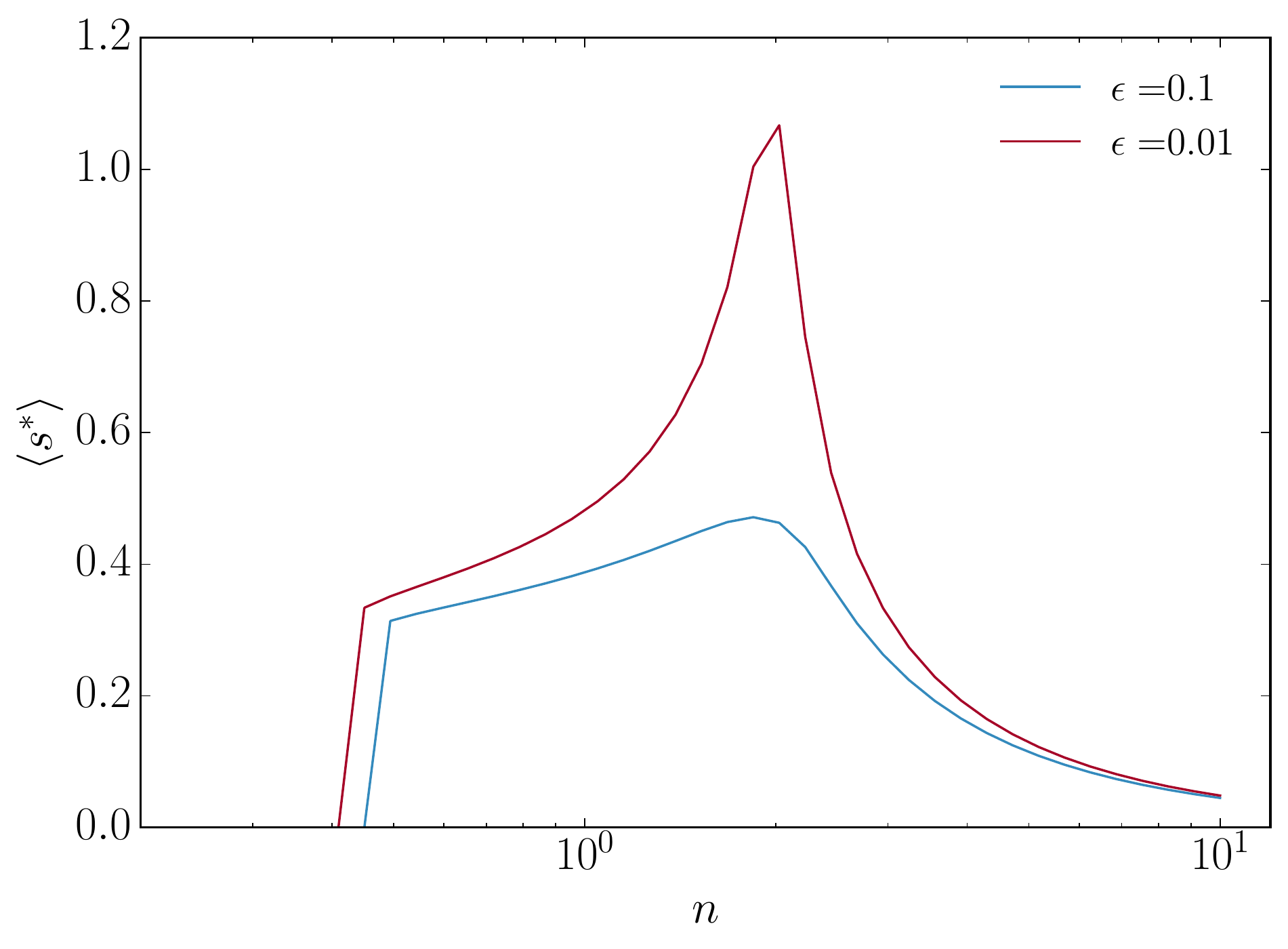}
\caption{\textbf{The role of the efficiency.} Comparison between the optimal scales of production $\langle s^* \rangle$, as functions of $n$, under different levels of inefficiency $\epsilon$ in the economy. Both lines have been obtained for $\pi = 0.65$ and $f = 0.5$.}
\label{eps_peak}
\end{figure}


\section{Consumption and waste} \label{sec:sm_cons}
In this Appendix we provide further insight about the behavior of the quantities introduced in equation (\ref{x_op}), from which the average consumption $X_C$ and the average waste $X_W$ can be defined (see equation \ref{CW}).

When no technologies are operating, only primary goods can be consumed or wasted, so the above quantities simplify to the following expressions:
\begin{eqnarray} \label{x_non_op}
X_C(s=0) &=& f \pi \\ \nonumber
X_W(s=0) &=& (1-f) \pi \ .
\end{eqnarray}

By comparing the expressions in equations (\ref{CW}) and (\ref{x_non_op}), one can see that for consumption and waste to be continuous one would need the following conditions to be satisfied upon approaching the transition (from within the operating phase): $x_{11}, x_{10} \rightarrow 1$, and $x_{01}, x_{00} \rightarrow 0$. By direct inspection of figure \ref{X}, one can see that these conditions are \emph{not} satisfied, i.e.\ $x_{11}, x_{10} < 1$ and $x_{01}, x_{00} > 0$ at the transition (the left endpoint of the curves). This is due to the discontinuous nature of the transition in $\langle s^* \rangle$ (see 
section \ref{sec:geometry} 
and \ref{sec:sm_geometry}): technologies begin to operate at strictly positive scales of productions, which implies that strictly positive amounts of non-primary goods will suddenly appear (hence $x_{01}, x_{00} > 0$) through the consumption of non-zero amounts of primary goods (hence $x_{11}, x_{10} < 1$).
\begin{figure}[t]
\centering
\includegraphics[width=0.48\columnwidth]{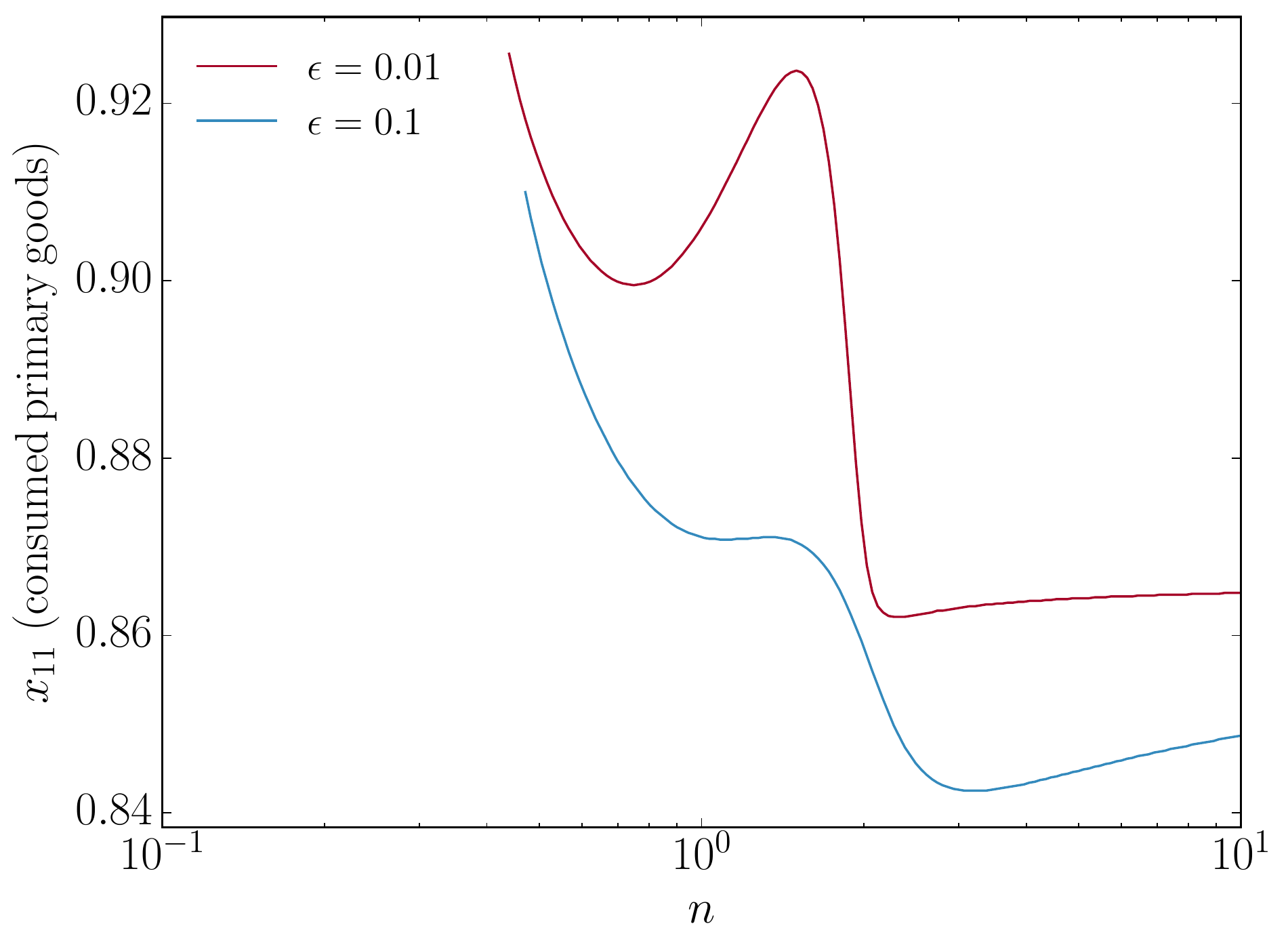}
\includegraphics[width=0.48\columnwidth]{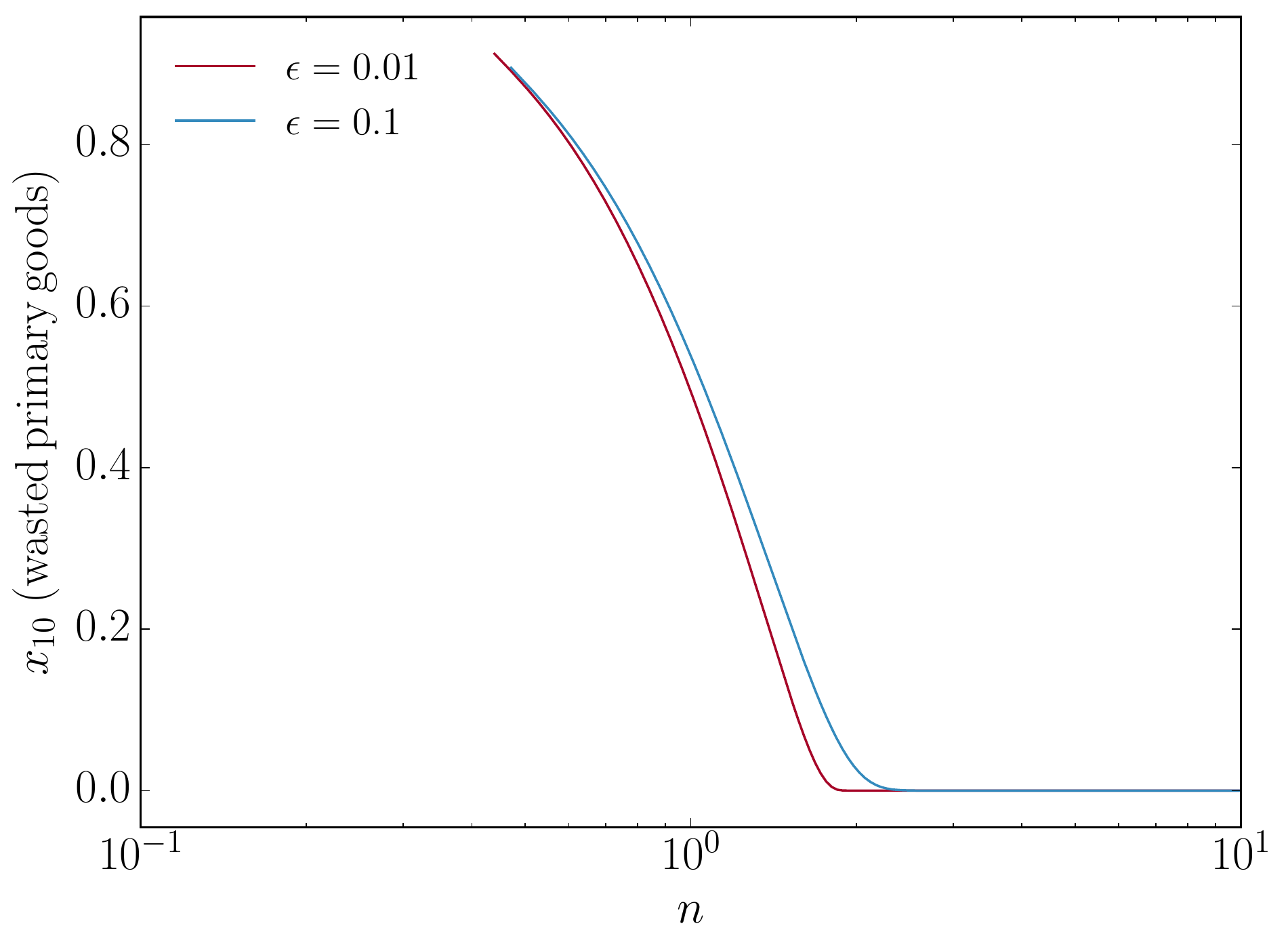}
\includegraphics[width=0.48\columnwidth]{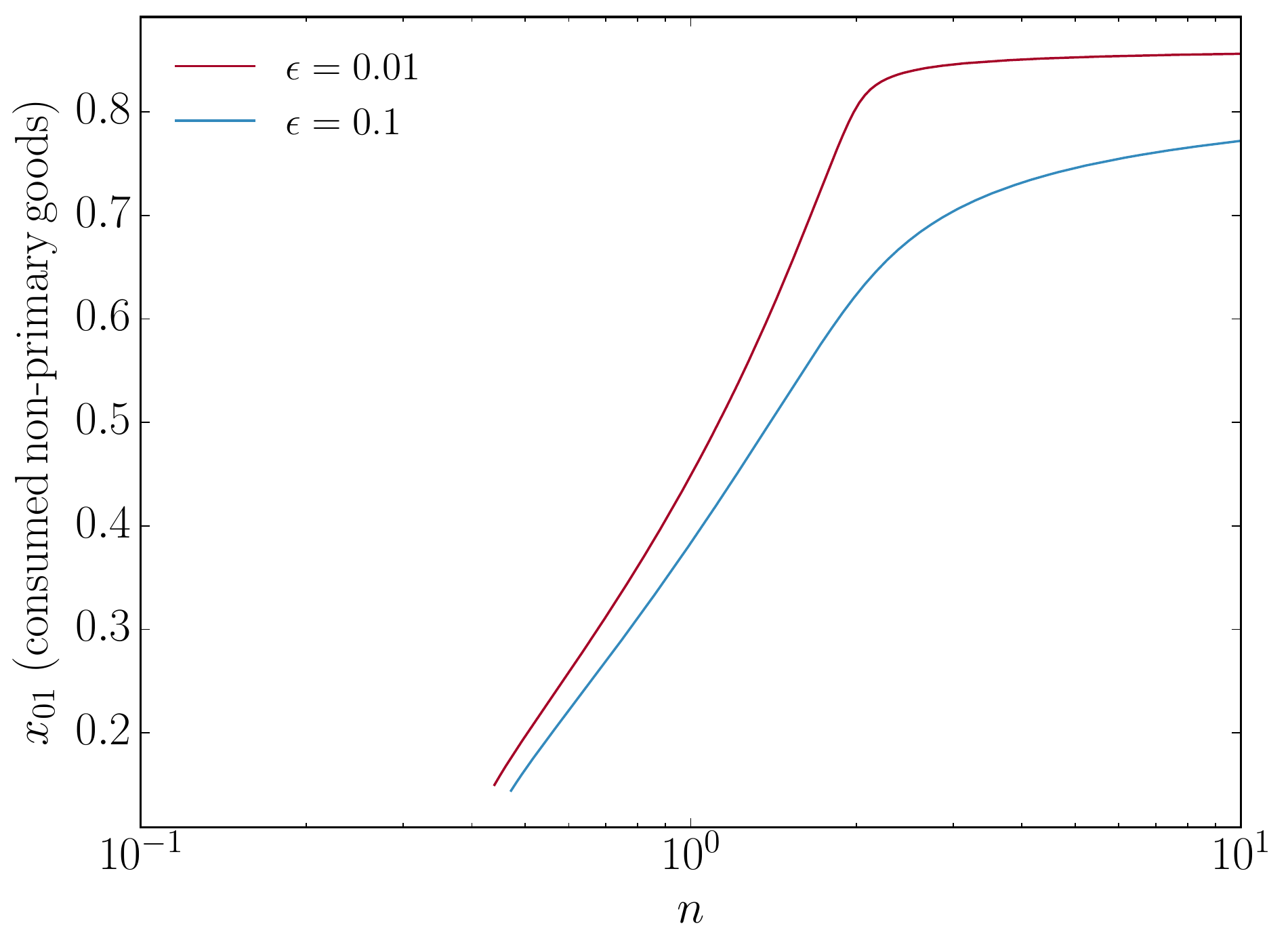}
\includegraphics[width=0.48\columnwidth]{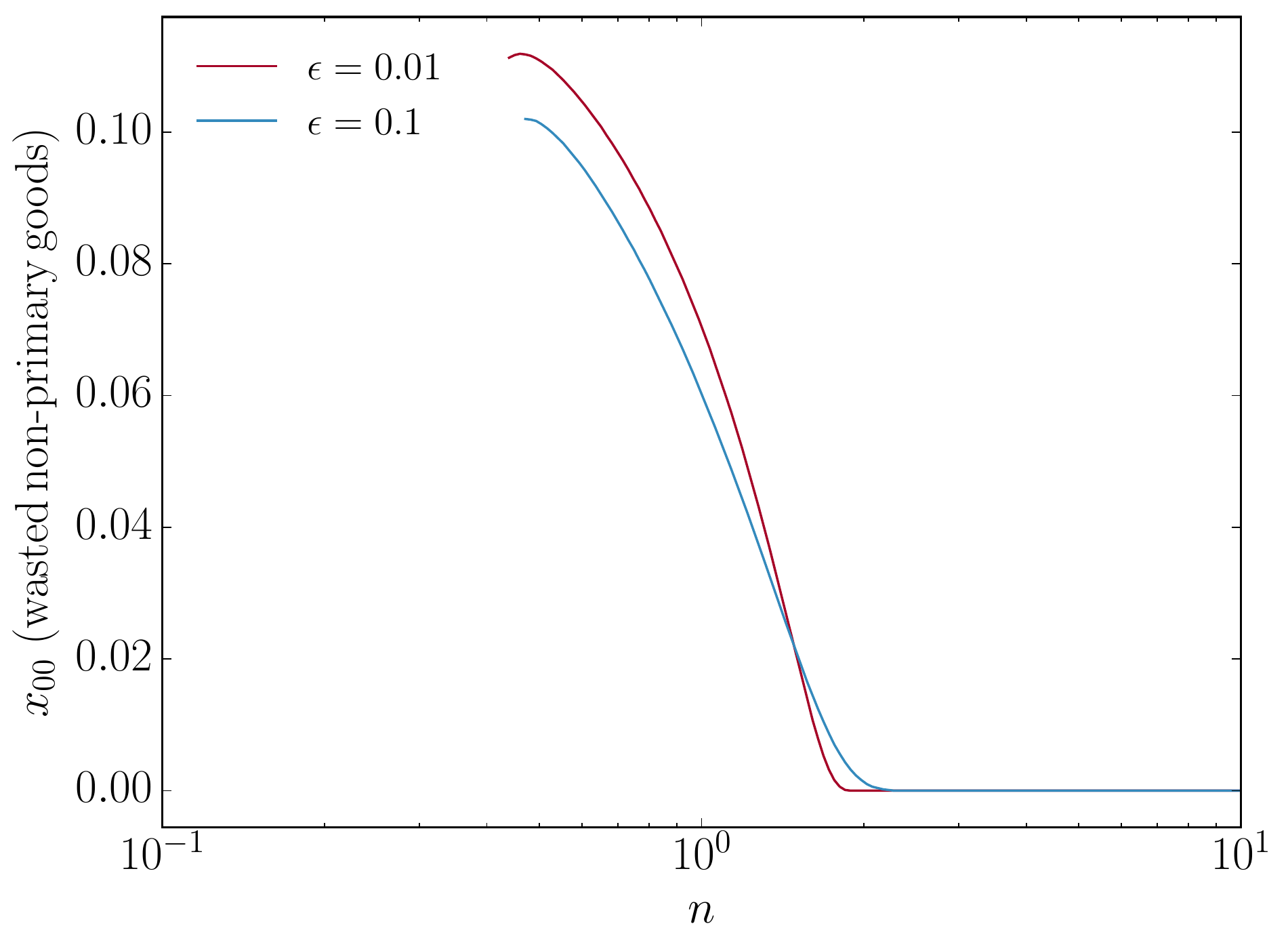}
\caption{\textbf{Consumption for all classes of goods.} Clockwise: behavior of $x_{11}$, $x_{10}$, $x_{01}$, $x_{00}$ in economies with $\pi = 0.65$, $f = 0.75$, and $\epsilon = 0.01$ (red curves) and $\epsilon = 0.1$ (blue curves).}
\label{X}
\end{figure}

Upon aggregation, the behaviors shown in figure \ref{X} cause the discontinuous behavior of both the consumption and the waste at the transition, shown in 
figure \ref{fig:wc_psi}. 
More efficient economies have higher average consumption levels, and a lower waste of primary goods. From figure \ref{X} we see that the disaggregated behaviour can be less trivial: consumed primary goods $x_{11}$ are not a monotonously increasing function of $n$ and more efficient economies do not necessarily waste less non-primary goods $x_{00}$ than less efficient ones (the curves for different values of $\epsilon$ cross in the bottom right panel).

The discontinuous behavior in $X_C$ and $X_W$ is reflected by the jump observed in 
$\langle x^* \rangle$. In the non-operational phase of the economy (i.e.\ where $s_i = 0 \ \forall i$), the production of goods becomes trivially identical to the initially available endowments: $x^c = x_0^c \ \forall c$, so that in this phase $\langle x^* \rangle_{s=0} = \pi$. When technologies are active one has $\langle x^* \rangle = \pi - n \epsilon \langle s^* \rangle$, so unless the economy is fully efficient ($\epsilon = 0$) the average production is not continuous at the onset of economic production (as shown in figure \ref{X_plot}). 
\begin{figure}[t]
\centering
\includegraphics[width=0.48\columnwidth]{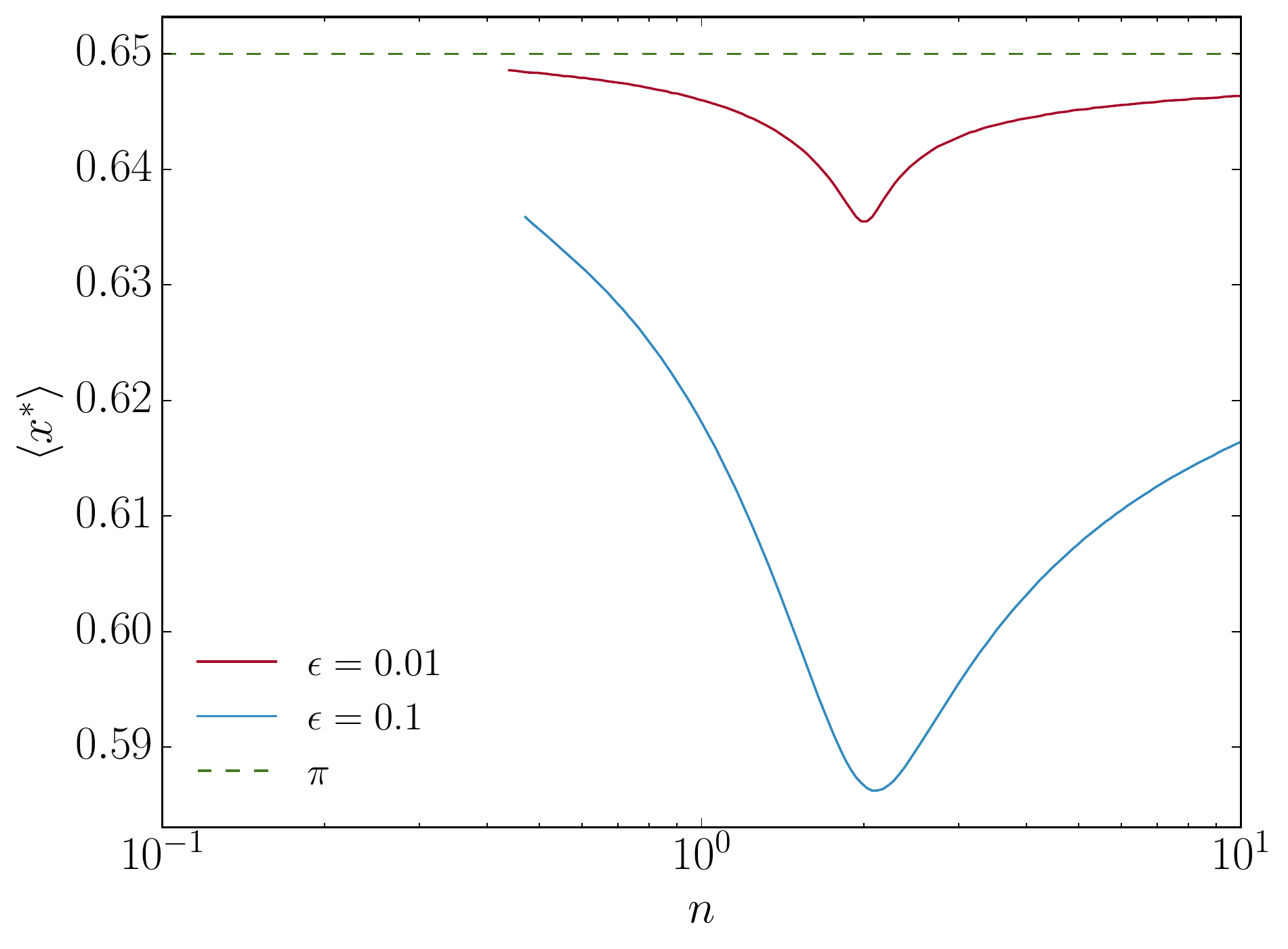}
\caption{\textbf{Aggregate consumption.} Behavior of $\langle x^* \rangle$ in an economy with $\pi = 0.65$, $f = 0.75$, and $\epsilon = 0.1$.}
\label{X_plot}
\end{figure}
All in all, at the transition there is a jump $\delta X = \langle x^* \rangle_{s=0} - \langle x^* \rangle = n \epsilon \langle s^* \rangle$. Clearly, $\delta X$ can be broken down into the corresponding jumps in consumption and waste, i.e.\ $\delta X = \delta X_C + \delta X_W$, where
\begin{eqnarray}
\delta X_C &=& X_C(s=0) - C = f \left [ \pi (1-x_{11}) - (1-\pi)x_{01} \right ] \\ \nonumber
\delta X_W &=& X_W(s=0) - W = (1-f) \left [ \pi (1-x_{10}) - (1-\pi) x_{00} \right ] \ .
\end{eqnarray}
More efficient economies correspond to a smaller $\delta X$, which in turn corresponds to smaller values of $\delta X_C$ and $\delta X_W$ 
(see figure \ref{fig:wc_psi}). 
Interestingly, it can be verified that a fully efficient economy where $\delta X = 0$ still displays discontinuous behavior in $\delta X_C$ and $\delta X_W$ (i.e.\ the continuous behavior in $\langle x^* \rangle$ is realized by having $\delta X_C = - \delta X_W$ rather than $\delta X_C = \delta X_W = 0$).


\section{Computing the Volume} \label{sec:sm_volume}
The aim of this section is to illustrate how the critical line separating the operational and non-operational phases can be computed analytically. In 
section \ref{sec:geometry}, 
we anticipate that the critical line corresponds to the vanishing of the volume $V$ defined by the $C$ constraints associated with goods in the $N$-dimensional space of scales of production. In the same section we also classify constraints as either non-homogeneous, which are associated with primary goods ($x_0^c \neq 0$) or homogeneous, which are associated with non-primary goods ($x_0^c = 0$). Homogeneous constraints are the only ones responsible for the vanishing of the volume and therefore, for the purpose of computing the critical line, it will suffice to compute the volume $V^\prime$ defined only by such constraints:
\begin{eqnarray} \label{eq:volume_true}
	V^\prime &&= \int_0^\infty \mathrm{d}\boldsymbol{s} \prod_{c \in \overline{\mathcal{P}}} \Theta \left( x_0^c + \sum_{i=1}^N q_i^c s_i \right) \\ \nonumber
	  &&= \int_0^\infty \mathrm{d}\boldsymbol{s} \prod_{c \in \overline{\mathcal{P}}} \Theta \left(\sum_{i=1}^N q_i^c s_i \right) \, ,
\end{eqnarray}
where the Heaviside $\Theta$ function ($\Theta(x) = 1$, for $x > 0$, and $\Theta(x) = 0$ otherwise) selects the region of the $N$-dimensional space compatible with the given constraints, and the product has been restricted to $\overline{\mathcal{P}}$, the set of non-primary goods.

The volume in equation (\ref{eq:volume_true}) depends on the specific realisation of technologies $\{q_i^c\}_{i=1, \ldots, N}^{c=1, \ldots, C}$. Since we are interested in the properties of large economies, we will average over the distribution of technologies and seek for self-averaging quantities. As pointed out in \cite{Marsili,BardosciaLivanMarsili}, the logarithm of the volume, and not the volume itself, is found to be self-averaging:
\begin{equation} \label{eq:h}
	h \equiv \lim_{N \rightarrow \infty} \frac{1}{N} \langle \log V^\prime \rangle_{\boldsymbol{q}, \boldsymbol{x}_0} \, .
\end{equation}
In fact, the limit of a vanishingly small volume corresponds to the limit $\chi \to 0$, where $\chi = \sum_{i=1}^N (s_{i,a} - s_{i,b})^2 / (2N)$ is the distance between different solutions (replicas) compatible with the constraints~\footnote{Let us explicitly point out that the quantity just introduced is slightly different from $\chi$ defined in equation (\ref{chi}). However, to keep consistency with \cite{Marsili} we prefer not to introduce another symbol.}. In this limit, the volume can be thought of as a tiny hypercube of side $\chi$, i.e.\ $V^\prime \simeq \chi^N = e^{N \log \chi}$. Hence, the quantity introduced in equation (\ref{eq:h}) must scale as $h \simeq \log \chi$.

Closely following the steps as in \cite{Marsili} (in particular see sections A and A.2 therein) one finds:
\begin{equation}
	h = g_1 + g_2 + g_3 \, ,
\end{equation}
where:
\begin{eqnarray} \label{eq:gs_bis}
	g_1 &=& \frac{\chi}{2} \left( \nu - \sigma^2 - \rho^2 \right) + \frac{1}{2} \nu \omega + \rho \lambda \, , \\ \nonumber
	g_2 &=& \Big\langle \log \int_0^\infty \mathrm{d}s \; e^{-\frac{\nu}{2} s^2 + \left[ \sigma t - \rho \epsilon \right] s} \Big\rangle_t \, , \\ \nonumber
	g_3 &=& \frac{1-\pi}{n} \Bigg\langle \log \frac{1}{2} \mathrm{Erfc}\left[ \frac{\sqrt{n \omega} t + n \lambda}{\sqrt{2 n \chi}} \right]\Bigg\rangle_t \, .
\end{eqnarray}
The variables $\chi$, $\nu$, $\sigma$, $\rho$, $\omega$, and $\lambda$ are called \emph{order parameters} and their value is set self-consistently by means of the \emph{saddle-point} equations:
\begin{equation} \label{eq:sp}
\frac{\partial h}{\partial \chi} = \frac{\partial h}{\partial \nu} = \frac{\partial h}{\partial \sigma} = \frac{\partial h}{\partial \rho} = \frac{\partial h}{\partial \omega} = \frac{\partial h}{\partial \lambda} = 0 \, .
\end{equation}
From equation (\ref{eq:sp}) it is easy to show that, while $\omega$ and $\lambda$ stay finite for $\chi \to 0$, $\nu$, $\sigma$, and $\rho$ scale like $1/\chi$. As a consequence, introducing
\begin{equation}
	\nu = \frac{v}{\chi} \qquad \sigma = \frac{c}{\chi} \qquad \rho = \frac{r}{\chi} \, ,
\end{equation}
and computing $\tilde{h} = \lim_{\chi \rightarrow 0} \chi h$, we ensure that all terms in $h$ diverging faster than $\log \chi$ in the limit $\chi \to 0$ are suppressed. By doing this we find $\tilde{h} = \tilde{h}_1 + \tilde{h}_2 + \tilde{h}_3$, where:
\begin{eqnarray} \label{eq:hs}
	\tilde{h}_1 &=& \frac{1}{2} \left( v \omega - c^2 - r^2 \right) + r \lambda \, , \\ \nonumber
	\tilde{h}_2 &=& \Big\langle \max_{s \geq 0} \left[ -\frac{v}{2} s^2 + \left( c t - r \epsilon \right) s \right] \Big\rangle_t \, , \\ \nonumber
	\tilde{h}_3 &=& - \frac{(1-\pi) \omega}{2 n} \langle \Theta(t + t_0) (t + t_0)^2 \rangle_t \, ,
\end{eqnarray}
with $t_0 = \sqrt{\frac{n}{\omega}} \lambda$. Exploiting the saddle point equations on $r$, $c$ and $v$ we can rewrite $\tilde{h}$ as a function of only two variables:
\begin{equation}
	\tilde{h} = \frac{c^2}{2} \left[ 1 + \frac{\xi^2}{\epsilon^2} - \frac{1-\pi}{n} \frac{I_2(-\xi)}{I_0(-\xi)^2} I_2(t_0) \right] \, ,
\end{equation}
with
\begin{eqnarray}
	t_0 &=&  \sqrt{\frac{n}{I_2(-\xi)}} \left[ \frac{\xi I_0(-\xi)}{\epsilon} +\epsilon I_1(-\xi)\right] \, , \\ \nonumber
	I_n(x) &=& \langle \Theta(t + x) (t + x)^n \rangle_t \, .
\end{eqnarray}
For a fixed value of $n$ we can now solve the two equations $\frac{\partial \tilde{h}}{\partial c} = 0$, $\frac{\partial \tilde{h}}{\partial \xi} = 0$ to find the critical value of $\pi$ at which the volume shrinks to zero. Therefore, as we vary the value of $n$, we are able to draw the critical line in the plane $(n, \pi)$. 

In order to check that the analytically computed critical line is correct we proceed as follows. The volume in equation (\ref{eq:volume_true}) is delimited by $N$-dimensional hyperplanes that either select an infinite region or a region of zero volume (the origin) in the space of scales of production. In the latter case the maximum of any linear combination of scale of productions $\{s_i\}_{i=1,\ldots,N}$ will be precisely $\boldsymbol{s} = 0$. Finding the maximum of any such function of scale of productions compatible with linear constraints is the very definition of a linear programming problem. Hence, by sampling from the distribution of technologies $\boldsymbol{q}$, one can solve different instances of the corresponding linear programming problem and count the fraction of instances that admit a solution other than $\boldsymbol{s} = 0$. In figure \ref{fig:vol} we show the behaviour of such quantity in the plane $(n, \pi)$. We can clearly distinguish two regions: in the lower left corner (blue region) none of the instances admits a non-trivial solution, while in the upper right corner (red region) all instances do. These two regions are separated by an intermediate region in which only some of the instances admit a non-trivial solution. However, as the size $N$ of the linear programming problem grows larger such intermediate region shrinks and the transition becomes sharper and sharper. From figure \ref{fig:vol} we can see that the analytically computed critical line sits in the middle of the transition region, and therefore is in excellent agreement with the numerical results. 

\begin{figure}
\centering
\includegraphics[width=0.48\columnwidth]{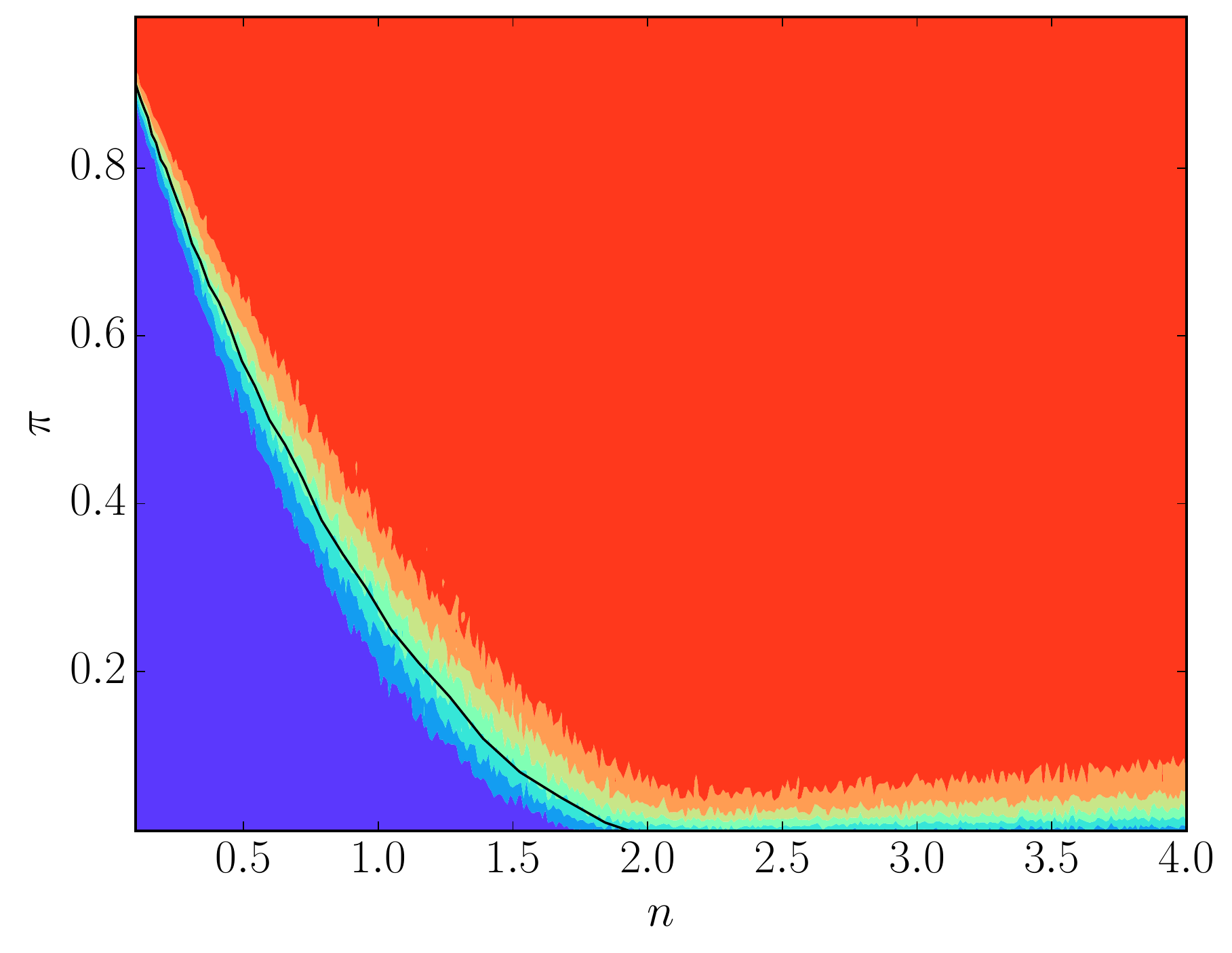}
\includegraphics[width=0.48\columnwidth]{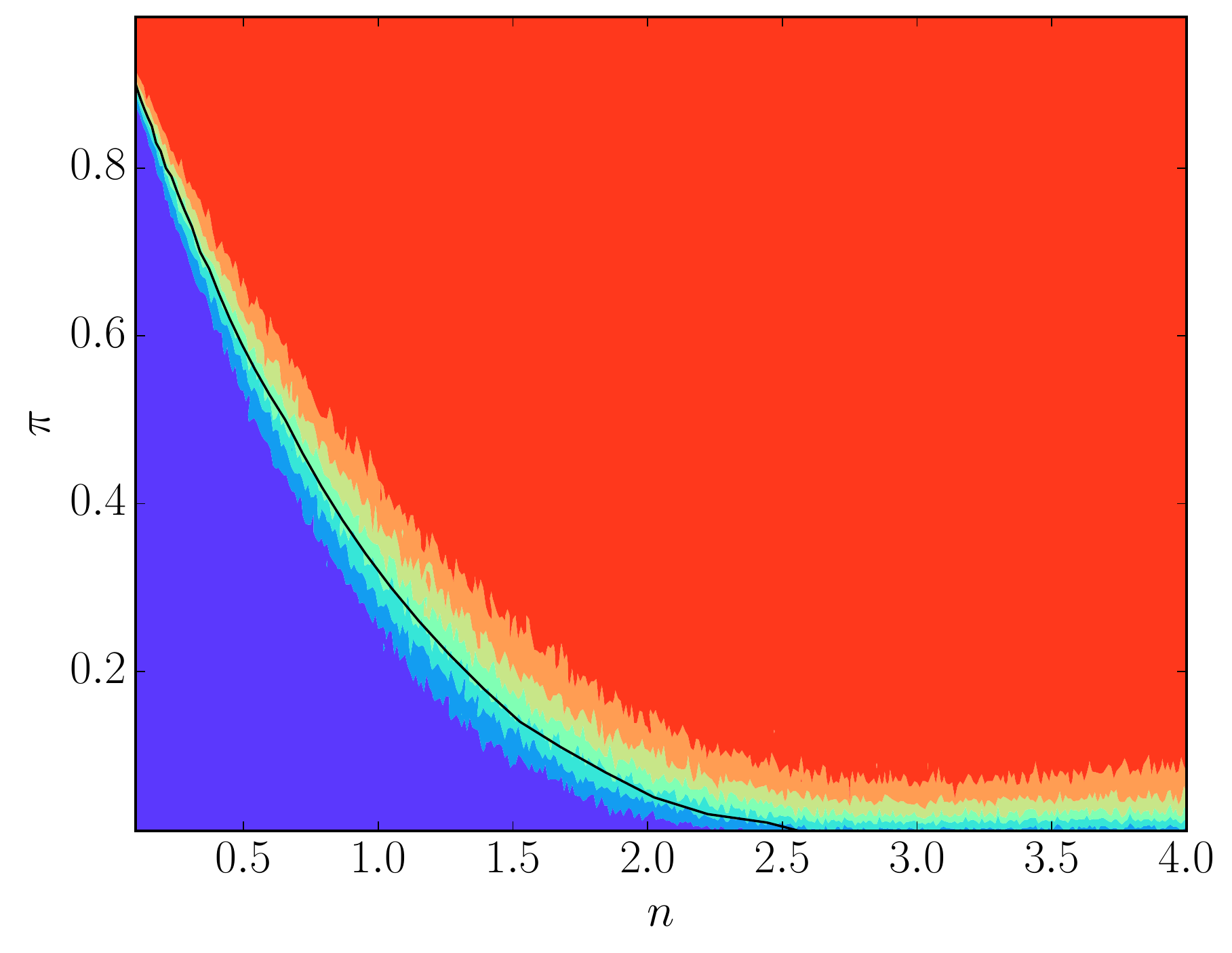}
\caption{\textbf{Numerical verification of the phase transition.} Fraction of (100) linear programming instances that admit a solution. In the blue region no instance admits a solution, while in red region all instances do. Linear programming instances have a finite size ($N = 100$) and therefore the transition between the operational phase and the non-operational phase has a finite width that shrinks as $N$ grows larger. The critical line computed analytically (solid black lines) is in excellent agreement with the linear programming numerics. $\epsilon = 0.01$ (left panel) and $\epsilon = 0.1$ (right panel).}
\label{fig:vol}
\end{figure}


\section{Geometric properties of the transition} \label{sec:sm_geometry}
In the previous section we have focused on computing the couples $(n, \pi)$ where the volume vanishes. Additional properties of the transition depend on \emph{how} the volume shrinks to zero. In particular, such transition can be either continuous or discontinuous. In order to characterise such behaviour we need to take into account the ``full'' volume: 
\begin{equation} \label{eq:volume_full}
	V = \int_0^\infty \mathrm{d}\boldsymbol{s} \prod_{c=1}^C \Theta \left( x_0^c + \sum_{i=1}^N q_i^c s_i \right) \, ,
\end{equation}
defined both by homogeneous and by non-homogeneous constraints. In fact, the transition for the volume in equation (\ref{eq:volume_true}) can only be discontinuous, since, for a given realization of the technologies, $V^\prime$ is either zero or infinite. Unfortunately, computing $V$ analytically is remarkably more difficult. However, solving the corresponding linear programming instances (for large $N$) is computationally feasible. Since the solutions of a linear programming problem lie on the boundary of the polytope identified by the problem's linear constraints, the surface of the volume $V$ (the feasible set) associated with a given realization of the technologies can be probed by fixing the problem's constraints (i.e.\ the input-output matrix $\boldsymbol{q}$) and optimizing multiple random linear functions (i.e.\ linear combinations of the scales of productions with random coefficients). Repeating this procedure for multiple (random) choices of the constraints provides access to the average properties of the volume $V$. As outlined in  
section \ref{sec:geometry}, 
we build the correlation matrix of the sampled solution vectors $\boldsymbol{s}^*$ and compute its largest eigenvalue $\lambda_{\mathrm{max}}$. Principal Components Analysis (PCA) informs us that $\lambda_{\mathrm{max}}$ will be close to $N$ whenever the volume $V$ has an elongated shape around a dominant direction. In figure \ref{fig:pca} we repeatedly compute $\lambda_{\mathrm{max}}$ for a fixed value of $n$ and for decreasing values of $\pi$, i.e.\ by ``approaching'' the transition by moving along what would be vertical lines in figure \ref{fig:vol}. We can see that $\lambda_{\mathrm{max}}$ gradually increases, approaching $N$ for $\pi$ approaching its critical value (the leftmost point in figure \ref{fig:pca}). Such behaviour is consistent with volumes $V$ gradually acquiring a dominant direction close to the transition, consistently with the sketch in figure \ref{fig:pictorial} of the paper.

\begin{figure}
\centering
\includegraphics[width=0.48\columnwidth]{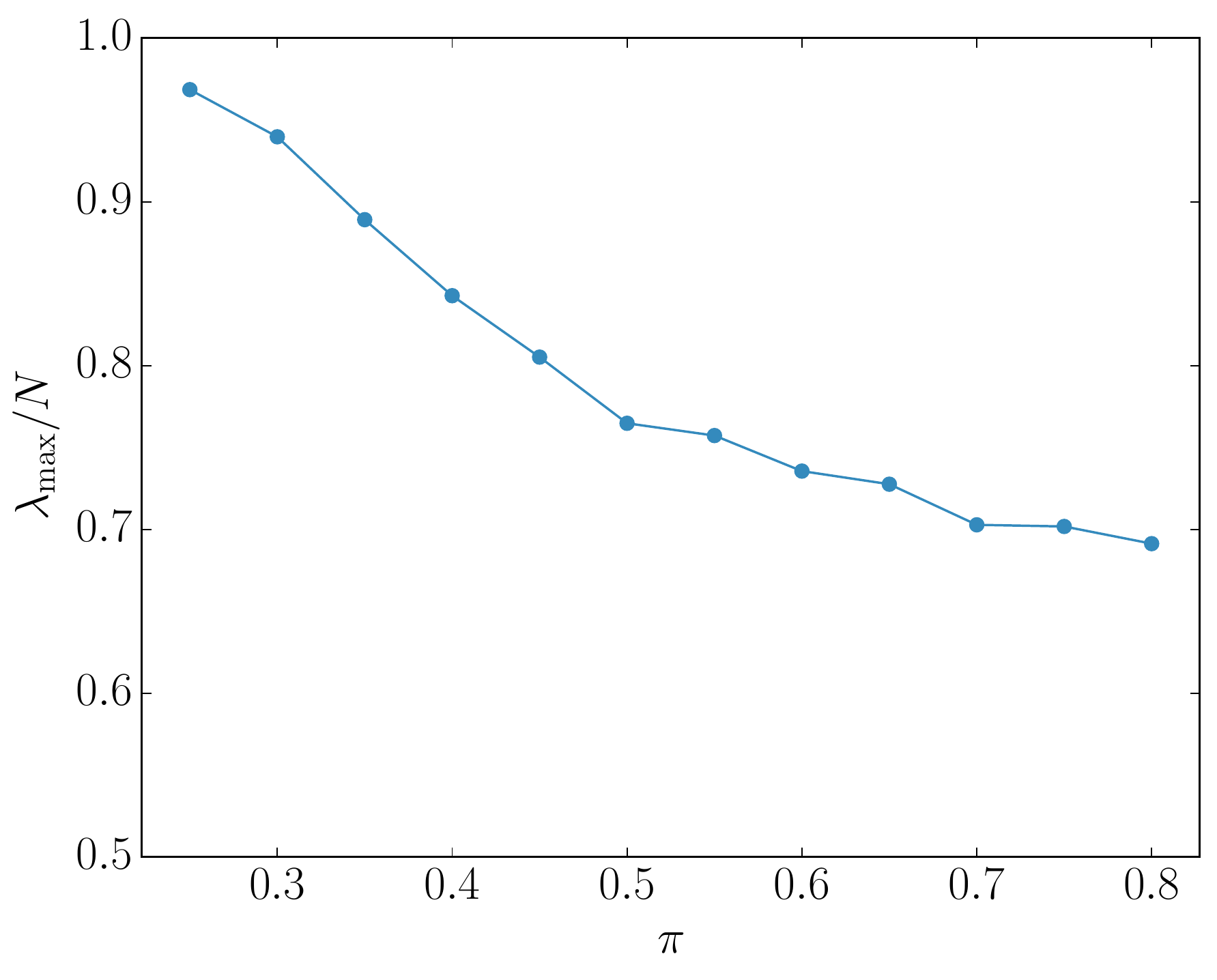}
\caption{\textbf{Sampling of the feasible set.} Principal component analysis on sampled solution vectors $\boldsymbol{s}^*$. Every point is averaged over 250 realizations (10 realizations of technologies, and for each realization of technologies 25 realizations of initial endowments and of random linear functions). Moving towards the critical line, i.e.\ towards smaller values of $\pi$, $\lambda_{\mathrm{max}}$ approaches $N$, showing that the feasible set has an elongated shape. $\epsilon = 0.01$, $N = 100$, $C = 100$ ($n = 1$).}
\label{fig:pca}
\end{figure}

\end{document}